\documentclass[12pt]{article}
\usepackage{amssymb}
\usepackage{amsmath}
\usepackage{amsthm}
\usepackage{latexsym}
\usepackage{mathdots}
\usepackage{imakeidx}
\makeindex
\newtheorem{theo}{Theorem}[section]
\newtheorem{prop}[theo]{Proposition}
\newtheorem{cor}[theo]{Corollary}

\theoremstyle{definition}
\newtheorem{defi}[theo]{Definition}
\newtheorem{exa}[theo]{Example}
\newtheorem{rem}[theo]{Remark}
\numberwithin{equation}{section}

\newcommand{\N}{{\mathbb N}}
\newcommand{\F}{{\mathbb F}}

\newcommand{\Q}{{\mathbb Q}}
\newcommand{\C}{{\mathbb C}}
\newcommand{\R}{{\mathbb R}}

\newcommand{\cC}{{\mathcal C}}

\newcommand{\cE}{{\mathcal E}}

\newcommand{\cM}{{\mathcal M}}

\newcommand{\cL}{{\mathcal L}}

\newcommand{\cR}{{\mathcal R}}

\newcommand{\GL}{\mbox{\rm GL}}
\newcommand{\Fxs}{\mbox{$\F[x;\sigma]$}}
\newcommand{\FFxs}{\mbox{$F[x;\sigma]$}}

\newcommand{\T}{\mbox{$^{\sf T}$}}
\newcommand{\bracket}[1]{[\![{#1}]\!]}
\newcommand{\ideal}[1]{\mbox{$({#1})$}}
\newcommand{\lideal}[1]{\mbox{$^{\bullet}({#1})$}}
\newcommand{\rideal}[1]{\mbox{$({#1})^{\bullet}$}}
\newcommand{\rs}{\mbox{\rm rs}}
\newcommand{\Aut}{\mbox{\rm Aut}}

\newcommand{\pp}{{\frak{p}}}
\newcommand{\vv}{{\frak{v}}}
\newcommand{\ov}[1]{\overline{#1}}

\newcommand{\rmid}{\mbox{$\,|_{r}\,$}}

\newcommand{\rank}{{\rm rk}}
\newcommand{\gcrd}{{\rm gcrd}}
\newcommand{\lclm}{{\rm lclm}}
\newcommand{\id}{{\rm id}}
\newcommand{\CircM}{{\rm Circ}}
\newcommand{\Mat}{{\rm Mat}}
\newcounter{alp}
\newcounter{ara}
\newcounter{rom}
\newenvironment{romanlist}{\begin{list}{(\roman{rom})\hfill}{\usecounter{rom}
     \topsep0ex \labelwidth.8cm \leftmargin.8cm \labelsep0cm
     \rightmargin0cm \parsep0ex \itemsep.4ex
     \partopsep1ex}}{\end{list}}
\newenvironment{alphalist}{\begin{list}{(\alph{alp})\hfill}{\usecounter{alp}
     \topsep-.4ex \labelwidth.7cm \leftmargin.7cm \labelsep0cm
     \rightmargin0cm \parsep0ex \itemsep.5ex
     \partopsep0ex}}{\end{list}}
\newenvironment{arabiclist}{\begin{list}{(\arabic{ara})\hfill}{\usecounter{ara}
     \topsep-.4ex \labelwidth.7cm \leftmargin.7cm \labelsep0cm
     \rightmargin0cm \parsep0ex \itemsep.3ex
     \partopsep0ex}}{\end{list}}

\begin{document}
\title{Introduction to Skew-Polynomial Rings\\[.6ex]  and Skew-Cyclic Codes\footnote{This survey will appear as a chapter in  ``A Concise Encyclopedia of Coding Theory'' to be published by CRC Press.}
}
\date{\today}
\author{Heide Gluesing-Luerssen\footnote{Partially supported by the grant \#422479 from the Simons Foundation.
  Department of Mathematics, University of Kentucky, Lexington KY 40506-0027, USA;
  heide.gl@uky.edu.}}

\maketitle

{\bf Abstract:}
This is a survey on the theory of skew-cyclic codes based on skew-polynomial rings of automorphism type.
Skew-polynomial rings have been introduced and discussed by Ore (1933).
Evaluation of skew polynomials and sets of (right) roots were first considered by Lam (1986) and studied in great detail by Lam and Leroy thereafter.
After a detailed presentation of the most relevant properties of skew polynomials, we survey the algebraic theory of skew-cyclic codes as introduced by
Boucher and Ulmer (2007) and studied by many authors thereafter.
A crucial role will be played by skew-circulant matrices.
Finally, skew-cyclic codes with designed minimum distance are discussed, and we report on two different kinds of skew-BCH codes, which were designed in 2014 and later.


\section{Introduction}\label{S-Intro}
In classical block coding theory, cyclic codes are the most studied class of linear codes with additional algebraic structure.
This additional structure turns out to be highly beneficial from a coding-theoretical point of view.
Not only does it allow the design of codes with large minimum distance, it also gives rise to very efficient algebraic decoding algorithms.
For further details we refer to \cite[Ch.~4 and~5]{HP03} in the textbook by Huffman/Pless and the vast literature on this topic.

Initiated by Boucher/Ulmer in \cite{BGU07,BoUl09a,BoUl09}, the notion of cyclicity has been generalized in various ways to skew-cyclicity
during the last decade.
In more precise terms the quotient space $\F[x]/(x^n-1)$, which is the ambient space for classical cyclic codes, is replaced by
$\Fxs/\lideal{x^n-1}$, where $\Fxs$ is the skew-polynomial ring induced by an automorphism~$\sigma$ of~$\F$ (see Definition~\ref{D-SkewPoly}),
and $\lideal{x^n-1}$ is the left ideal generated by $x^n-1$.
Further generalizations are obtained by replacing the modulus $x^n-1$ by $x^n-a$, leading to skew-constacyclic codes, or even more general polynomials~$f$ of degree~$n$.
In any case, the quotient is isomorphic as a left $\F$-vector space to $\F^n$, and thus we may consider linear codes in~$\F^n$ as subspaces of the quotient.

This allows us to define skew-cyclic codes.
A linear code in~$\F^n$ is  $(\sigma,f)$-skew-cyclic if it is a left submodule of $\Fxs/\lideal{f}$.
As in the classical case, every such submodule is generated by a right divisor of the modulus~$f$.
If~$f=x^n-a$ or even $f=x^n-1$, the resulting codes are called $(\sigma,a)$-skew-constacyclic or $\sigma$-skew-cyclic, respectively.
The first most striking difference to the classical case is that the modulus $x^n-1$ has in general far more right divisors
in $\Fxs$ than in $\F[x]$.
As a consequence, a polynomial may have more roots than its degree suggests.
All of this implies that the family of skew-cyclic codes of given length is far larger than that of cyclic codes.

While these basic definitions are straightforward, a detailed study of the algebraic and coding-theoretic properties of skew-cyclic codes requires an understanding of
the skew-poly\-no\-mi\-al ring $\Fxs$.
In Sections~\ref{S-Basics},~\ref{S-Roots}, and~\ref{S-AlgSets} we will present the theory of skew-polynomials as it is needed for our study of skew-cyclic codes.
This entails division  properties in the ring~$\Fxs$, evaluations of skew polynomials and their (right) roots, and algebraic sets along with a skew version of Vandermonde matrices.
Division and factorization properties were studied in detail by Ore, who introduced skew-polynomial rings in the 1930's in his seminal paper~\cite{Ore33}.
Evaluations of skew polynomials were first considered by Lam~\cite{Lam86} in the 1980's and then further investigated by Lam and Leroy, see
\cite{Lam86,LaLe88,LaLe04,LLO08,Ler12}.
For Sections~\ref{S-Basics},~\ref{S-Roots}, and~\ref{S-AlgSets}, we will closely follow these sources.
We will add plenty of examples illustrating the differences to commutative polynomial rings.
In Section~\ref{S-Linearized} we will briefly present the close relation between skew polynomials over a finite field and linearized polynomials, which
play a crucial role in the area of rank-metric codes.

The material in Sections~\ref{S-Basics},~\ref{S-Roots}, and~\ref{S-AlgSets} provides us with the right toolbox to study skew-cyclic codes and their generalizations.
In Sections~\ref{S-BSCC} and~\ref{S-SkewConsta} we will derive the algebraic theory of $(\sigma,f)$-skew-cyclic codes and specialize to skew-constacyclic codes whenever necessary.
We will do so by introducing skew circulant matrices because their row spaces are the codes in question.
As a guideline to skew circulants we give a brief approach to classical cyclic codes via classical circulants in Section~\ref{S-CBC}.
Among other things we will see in Section~\ref{S-SkewConsta} that the dual of a $(\sigma,x^n-a)$-skew-constacyclic code is $(\sigma,x^n-a^{-1})$-skew-constacyclic, and a generator polynomial arises as a certain kind of
reciprocal of the generator polynomial of the primal code.
This result appeared first in~\cite{BoUl09a} and was later derived in more conceptual terms in~\cite{FGL15}.

In Section~\ref{S-MinDist} we will report on constructions of skew-cyclic codes with designed minimum distance.
The results are taken from the very recent papers \cite{BoUl14,GLNN18,TCTi17}.
They amount to essentially two kinds of skew-BCH codes.
For the first kind the generator polynomial has right roots that appear as consecutive powers of a suitable element in a field extension (similar to classical BCH codes),
whereas for the second kind it has right roots that are consecutive Frobenius powers of a certain element.
Both cases can be generalized to Hartmann-Tzeng form as for classical cyclic codes.
The theory of skew Vandermonde matrices will be a natural tool in the discussion.

We wish to stress that in this survey we restrict ourselves to skew-cyclic codes derived from
skew polynomials of automorphism type over fields.
More general situations have been studied in the literature; see Remark~\ref{R-GenSkew}.
In addition, some results have been obtained for quasi-skew-cyclic codes. We will not survey on this material either.
Finally, we also do not discuss decoding algorithms for the codes of Section~\ref{S-MinDist} in this survey.

\section{Basic Properties of Skew-Polynomial Rings}\label{S-Basics}
In this section we introduce skew-polynomial rings with coefficients in a field.
These rings were considered and studied first by Ore in~\cite{Ore33}.
We will give a brief account of the ring-theoretic results from~\cite{Ore33} insofar as they are important for our later discussions of skew-cyclic codes.

\begin{defi}\label{D-SkewPoly}
Let $F$ be any field and $\sigma\in\Aut(F)$.
\index{Skew-polynomial ring}
The  \textbf{skew-polynomial ring} $\FFxs$ is defined as the set
$\big\{\sum_{i=0}^N f_i x^i\,\big|\, N\in\N_0,f_i\in F\big\}$ endowed with usual addition, i.e., coefficientwise,
and multiplication given by the rule
\begin{equation}\label{e-xa}
   xa=\sigma(a)x\text{ for all }a\in F
\end{equation}
along with distributivity and associativity.
Then $(\FFxs,+,\,\cdot\,)$ is a ring with identity $x^0=1$.
Its elements are called \textbf{skew polynomials}\index{Skew polynomial} or simply polynomials.
\end{defi}

If $\sigma=\id$, then $\FFxs=F[x]$, the classical commutative polynomial ring over~$F$.
We refer to this special case as the \emph{commutative case} and \emph{commutative polynomials}.
In the general case, the additive groups of $\FFxs$ and $F[x]$ are identical, whereas
multiplication in $\FFxs$ is given by
\[
   \Big(\sum_{i=0}^N f_i x^i\Big)\Big(\sum_{j=0}^M g_j x^j\Big)=\sum_{i,j}f_i\sigma^i(g_j)x^{i+j}.
\]
Note that the set of skew polynomials may also be written as $\{\sum_{i=0}^N x^i f_i\mid N\in\N_0,f_i\in F\}$, i.e., with coefficients on the right of~$x$.
The only rule we have to obey is to apply $\sigma$ when moving coefficients from the right to the left of~$x$,
and thus $\sigma^{-1}$ for the other direction.
We will always write polynomials as $\sum_{i=0}^N f_i x^i$, and coefficients are meant to be left coefficients.
As a consequence, the \textbf{leading coefficient}\index{Skew polynomial!Leading coefficient}  of a polynomial is meant to be
its left  leading coefficient.

Note that $\FFxs$ is a left and right vector space over~$F$, but these two vector space structures are not identical.

\begin{rem}\label{R-GenSkew}
Skew polynomial rings are commonly introduced and studied in much more generality.
One may replace the coefficient field~$F$ by a division algebra or even a noncommutative ring; one may
consider (ring) endomorphisms~$\sigma$ instead of automorphisms; and one may introduce a $\sigma$-derivation, say~$\delta$, which then turns~\eqref{e-xa} into
$xa=\sigma(a)x+\delta(a)$.
All of this is standard in the literature of skew-polynomial rings.
For simplicity of this presentation we restrict ourselves to skew-polynomial rings as in Definition~\ref{D-SkewPoly}.
However, we wish to point to the article~\cite{BoUl14}  for examples showing that a $\sigma$-derivation may indeed lead to skew-cyclic codes with better minimum
distance than what can be achieved with the aid of an automorphism alone.
In addition, we will not discuss skew-polynomial rings with coefficients from a finite ring.
\end{rem}

\begin{rem}\label{R-Center}
Consider the skew-polynomial ring $\FFxs$, and let $K\subseteq F$ be the fixed field of~$\sigma$.
If~$\sigma$ has finite order, say~$m$,  the center of $\FFxs$ is given by the commutative polynomial ring $K[x^m]$.
This is easily seen by using the fact that any~$f$ in the center satisfies $xf=fx$ and $af=fa$ for all $a\in F$.
If~$\sigma$ has infinite order, the center is~$K$.
\end{rem}

\begin{exa}\label{E-Complex}
Consider the field~$\C$ of complex numbers, and let~$\sigma$ be the complex conjugation.
Then the center of $\C[x;\sigma]$ is the commutative polynomial ring $\R[x^2]$.
Furthermore, $\R[x]$ is a subring of both $\C[x;\sigma]$ and $\C[x]$, which shows that a skew-polynomial ring may be a subring of skew-polynomial rings with different automorphisms.
\end{exa}

Later we will restrict ourselves to skew-polynomial rings over finite fields.
The following situation essentially covers all such cases because each automorphism is a power of the Frobe\-nius automorphism over the prime field.
Throughout, the automorphism $\F_{q^m}\rightarrow\F_{q^m}$ given by $c\mapsto c^q$ is simply called the \textbf{$q$-Frobenius}.\index{$q$-Frobenius}

\begin{exa}\label{E-FxsFinite}
Consider the skew-polynomial ring $\F[x;\sigma]$ where $\F=\F_{q^m}$ and $\sigma\in\Aut(\F)$ is the $q$-Frobenius.
Then~$\sigma$ has order~$m$ and fixed field~$\F_q$, and the center of $\F[x;\sigma]$ is $\F_q[x^m]$.
\end{exa}

Let us return to general skew-polynomial rings $\FFxs$ and fix the following standard notions.

\begin{defi}\label{D-Degree}
The \textbf{degree} of skew polynomials is defined in the usual way as the largest exponent of~$x$ appearing in the polynomial, and $\deg(0):=-\infty$.
This does not depend on the side where we place the coefficients because~$\sigma$ is an automorphism, and we obtain
\[
  \deg(f+g)\leq\max\{\deg(f),\deg(g)\}\ \text{ and }\
  \deg(fg)=\deg(f)+\deg(g).
\]
As a consequence, the group of units of $\FFxs$ is given by $F^*=F\setminus\{0\}$.
A nonzero polynomial is \textbf{monic} if its leading coefficient is~$1$. Again, this does not depend on the sidedness of the coefficients because $\sigma(1)=1$.
We say that~$g$ is a \textbf{right divisor}\index{Skew polynomial!Right divisor} of $f$ and write $g\rmid f$ if $f=hg$ for some $h\in \FFxs$.
A polynomial $f\in \FFxs\setminus F$ is \textbf{irreducible}\index{Skew polynomial!Irreducibility} if all its right (hence left) divisors are units or polynomials of the same degree as~$f$.
Clearly, polynomials of degree~$1$ are irreducible.
\end{defi}

\begin{exa}\label{E-SkewPolyF4}
Let $F=\F_4=\{0,1,\omega,\omega^2\}$, where $\omega^2=\omega+1$, and let~$\sigma$ be the 2-Frobenius.
Then $\sigma^{-1}=\sigma$.
In $\F_4[x;\sigma]$ we have
\begin{arabiclist}
\item $x^2+1=(x+1)(x+1)=(x+\omega^2)(x+\omega)=(x+\omega)(x+\omega^2)$.
         Thus, a skew polynomial may have more linear factors than its degree suggests.
\item $(x^2+\omega x+\omega)(x+\omega)=x^3+\omega^2x+\omega^2$, and thus $x+\omega$ is a right divisor of~$x^3+\omega^2x+\omega^2$.
        It is not a left divisor. This is easily seen by computing
        \[
          ( x+\omega)(f_2x^2+f_1x+f_0)=f_2^2x^3 +(\omega f_2+f_1^2)x^2+(\omega f_1+f_0^2)x+\omega f_0
        \]
        and comparing coefficients with those of $x^3+\omega^2x+\omega^2$.
\item The polynomial $x^{14}+1\in\F_4[x;\sigma]$ has $599$ nontrivial monic right divisors.
         On the other hand, in the commutative polynomial ring $\F_4[x]$ the same polynomial has only~$25$ nontrivial monic right divisors.
\end{arabiclist}
\end{exa}

Just as for commutative polynomials, one can carry out division with remainder in $\FFxs$ if one takes sidedness of the coefficients into account.
This is spelled out in~(a) below.
The proof  is entirely analogous to the commutative case.
Indeed, if $\deg(f)=m\geq\deg(g)=\ell$ and the leading coefficients of~$f$ and~$g$ are $f_m$ and $g_\ell$, respectively, then the polynomial
$f-f_m\sigma^{m-\ell}(g_\ell^{-1})x^{m-\ell}g$ has degree less than~$m$.
This allows one to proceed until a remainder of degree less than~$\ell$ is obtained.
The rest of the theorem formulates the familiar consequences of division with remainder.

\begin{theo}[\mbox{\cite[p.~483--486]{Ore33}}]\label{R-PropR}
$\FFxs$ is a left Euclidean domain and a right Euclidean domain.
More precisely, we have the following.
\begin{alphalist}
\item \textbf{Right division with remainder:}\index{Skew-polynomial ring!Right division with remainder} For all $f,\,g\in \FFxs$ with $g\neq0$ there exist unique polynomials $s,r\in \FFxs$ such that
      $f=sg+r$ and $\deg(r)<\deg(g)$.
      If $r=0$, then~$g$ is a right divisor of~$f$.
\item For any two polynomials $f_1,\,f_2\in\FFxs$, not both zero, there exists a unique monic polynomial~$d\in \FFxs$
      such that $d\rmid f_1,\ d\rmid f_2$ and such that whenever $h\in\FFxs$ satisfies
      $h\rmid f_1$ and $h\rmid f_2$ then $h\rmid d$.
      The polynomial~$d$ is called the \textbf{greatest common right divisor}\index{Skew-polynomial ring!Greatest common right divisor} of $f_1$ and~$f_2$, denoted by $\gcrd(f_1,f_2)$.
      It satisfies a \textbf{right Bezout identity}\index{Skew-polynomial ring!Right Bezout identity}, that is,
      \[
          d=uf_1+vf_2\ \text{ for some }u,\,v\in \FFxs.
      \]
      We may choose $u,\,v$ such that $\deg(u)<\deg(f_2)$ and, consequently, $\deg(v)<\deg(f_1)$.
      This is a consequence of the Euclidean algorithm, see also~\cite[Sec.~2]{Gie98}.
      If $d=1$, we call $f_1,f_2$ \textbf{relatively right-prime}.\index{Skew-polynomial ring!Relatively right-prime}
\item For any two nonzero polynomials $f_1,\,f_2\in \FFxs$, there exists a unique monic polynomial~$\ell\in \FFxs$
      such that $f_i\rmid\ell,\,i=1,2,$ and such that whenever $h\in \FFxs$ satisfies
      $f_i\rmid h,\,i=1,2,$ then $\ell\rmid h$.
      The polynomial~$\ell$ is called the \textbf{least common left multiple}\index{Skew-polynomial ring!Least common left multiple} of $f_1$ and~$f_2$, denoted by $\lclm(f_1,f_2)$.
      Moreover, $\ell=uf_1=vf_2$ for some $u,\,v\in \FFxs$ with $\deg(u)\leq\deg(f_2)$ and $\deg(v)\leq\deg(f_1)$.
\item For all nonzero $f_1,\,f_2\in \FFxs$
      \[
           \deg(\gcrd(f_1,f_2))+\deg(\lclm(f_1,f_2))=\deg(f_1)+\deg(f_2).
      \]
\end{alphalist}
Analogous statements hold true for the left hand side.
\end{theo}

We refer to~\cite{Gie98,CaBo17} for various algorithms for fast computations in $\F[x;\sigma]$ for a finite field~$\F$,
in particular for factoring skew polynomials into irreducibles.

Exactly as in the commutative case, the above leads to the following consequence.

\begin{theo}\label{T-PIR}
Let $I\subseteq \FFxs$ be a left ideal (that is,~$(I,+)$ is a subgroup of $(\FFxs,+)$ and~$I$ is closed with respect to left multiplication by elements
from $\FFxs$). Then~$I$ is principal, i.e., there exist $f\in I$ such that
$I=\FFxs f=\{gf\mid g\in\FFxs\}$.
For brevity we will use the notation $\lideal{f}$ for $\FFxs f$ and call it the \textbf{left ideal generated by}~$f$.\index{Skew-polynomial ring!Left principal ideal $\lideal{f}$}
An analogous statement is true for right ideals.
Thus, $\FFxs$ is a \textbf{left principal ideal ring} and a \textbf{right principal ideal ring}.
\end{theo}

Recall the center of $\FFxs$ from Remark~\ref{R-Center}.
It is clear that for any polynomial~$f$ in the center, the ideal $\lideal{f}$ is \textbf{two-sided}, \index{Skew-polynomial ring!Two-sided ideal}i.e.\ a left ideal and a right ideal.
Polynomials generating two-sided ideals are closely related to central elements.

\begin{rem}\label{R-TwoSided}
Let~$\sigma$ have order~$m$.
An element~$f\in \FFxs$ is called \textbf{two-sided}\index{Skew polynomial!Two-sided} if the ideal $\lideal{f}$ is two-sided, i.e.,
$\lideal{f}=\rideal{f}$.
It is not hard to see \cite[Thm.~1.1.22]{Jac96} that the two-sided elements of~$\FFxs$ are exactly the polynomials of the form
$\{c x^t g\mid c\in F,\,t\in\N_0, g\in Z\}$, where $Z=K[x^m]$ is the center of~$\FFxs$.
As a  special case, for any $a \in F^*$, the polynomial $x^n-a$ is two-sided if and only if it is central if and only if $m\mid n$ and $\sigma(a)=a$.
\end{rem}

With respect to many properties addressed thus far, the skew-polynomial ring $\FFxs$ behaves similar to the commutative polynomial ring $F[x]$.
However, a main difference is that in  $\FFxs$, where $\sigma\neq\id$, polynomials do not factor uniquely (up to order) into irreducible polynomials.
We have seen this already in Example~\ref{E-SkewPolyF4}(1).
Of course, irreducible factorizations of nonunits still exist, which is a consequence of the boundedness of the degree by zero.
In order to formulate a uniqueness result, we need the notion of similarity defined in~\cite{Ore33}. The equivalence of~(i) and~(ii) below is straightforward.
The equivalence to~(iii) can be found in~\cite[Prop.~1.2.8]{Jac96}.

\begin{defi}\label{D-similar}
Let $f,g\in\FFxs$. The following are equivalent.
\begin{romanlist}
\item There exist $h,\,k\in \FFxs$ such that $\gcrd(f,h)=1,\,\mbox{\rm gcld}(g,k)=1$,  and $gh=kf$.
\item There exist $h\in \FFxs$  such that $\gcrd(f,h)=1$ and $\lclm(f,h)= g  h$.
\item The left $\FFxs$-modules $\FFxs/\lideal{f}$ and $\FFxs/\lideal{g}$ are isomorphic.
\end{romanlist}
If~(i), hence~(ii) and~(iii), holds true, the polynomials $f,\,g$ are called (left) \textbf{similar}.\index{Skew-polynomial ring!Similarity}
\end{defi}

In the commutative ring~$F[x]$, two polynomials are thus similar if and only if they differ by a constant factor.
In general, similar polynomials have the same degree (see Proposition~\ref{R-PropR}(d)).
Part~(iii) shows that similarity is indeed an equivalence relation on $\FFxs$ and does not depend on the sidedness, i.e., $f,g$ are right similar iff they are left similar
(see also \cite[Thm.~13, p.~489]{Ore33} without resorting to~(iii)).
Part~(iii) above is the similarity notion for left ideals as introduced and discussed by Cohn \cite[Sec.~3.2]{Co85}.
It leads to a simple criterion for similarity, which we will present next.

For a monic polynomial $f=\sum_{i=0}^{n-1} f_ix^i+x^n\in\FFxs$ define the ordinary \textbf{companion matrix}\index{Companion matrix}
\begin{equation}\label{e-CompMat}
   C_f=\begin{pmatrix} &1&& &\\ & &\ddots& &\\ & & &1&\\ & & & &1\\-f_0&-f_1&\cdots&-f_{n-2}&-f_{n-1}\end{pmatrix}\in\Mat_{n,n}(F).
\end{equation}
Consider the map~$L_x$ given by left multiplication by~$x$ in the left $\FFxs$-module $\cM_f:=\FFxs/\lideal{f}$.
This map is $\sigma$-semilinear, i.e., $L_x(a t)=\sigma(a)L_x(t)$ for all $a\in F$ and $t\in\cM_f$.
Furthermore, the rows of $C_f$ are the coefficient vectors of $L_x(x^i)$ for $i=0,\ldots,n-1$.
All of this shows that $L_x(\sum_{i=0}^{n-1}a_i x^i)=\sum_{i=0}^{n-1}b_i x^i$, where
$(b_0,\ldots, b_{n-1})=(\sigma(a_0),\ldots,\sigma(a_{n-1}))C_f$.
In this sense,~$C_f$ is the matrix representation of the semi-linear map $L_x$ with respect to the basis $\{1,x,\ldots,x^{n-1}\}$.

If~$g$ is another monic polynomial of degree~$n$, then both $\cM_f$ and $\cM_g$ are $n$-dimensional over~$F$  and thus isomorphic $F$-vector spaces.
They are isomorphic as left $\FFxs$-modules if we can find a left $\FFxs$-linear isomorphism.
Along with Definition~\ref{D-similar}(iii) this easily leads to the following criterion for similarity (see also \cite[Thm.~4.9]{LLO08}).

\begin{prop}\label{P-CfCg}
Let $f,g\in\FFxs$ be monic polynomials of degree~$n$. Then $f,g$ are similar iff there exists a matrix $B\in\GL_n(F)$ such that
$C_g=\sigma(B)C_fB^{-1}$.
\end{prop}

In \cite[Prop.~2.1.17]{CaBo17} a different criterion is presented for finite fields~$F$ in terms of the reduced norm.

Now we are ready to formulate the uniqueness result for irreducible factorizations.

\begin{theo}[\mbox{\cite[p.~494]{Ore33}}]\label{T-IrredFac}
Let $f_1,\ldots,f_r,\,g_1,\ldots,g_s$ be irreducible polynomials in $\FFxs$ such that
$f_1\cdots f_r=g_1\cdots g_s$.
Then $r=s$ and there exists a permutation $\pi$ of $\{1,\ldots,r\}$ such that $g_{\pi(i)}$ is similar to~$f_i$ for all~$i=1,\ldots,r$.
In particular, $\deg(f_i)=\deg(g_{\pi(i)})$ for all~$i$.
\end{theo}

The reader should be aware that the converse of the above statement is not true: it is not hard to find (monic) polynomials $f_i,g_i,\,i=1,2$,
such that~$f_i$ and~$g_i$ are similar for $i=1,2$, but
$f_1f_2$ and $g_1g_2$ are not similar (and thus certainly not equal).

We close this section with the following example illustrating yet some further challenging features in factorizations of skew polynomials.

\begin{exa}\label{E-x21}
Consider $\F_4[x;\sigma]$ as in Example~\ref{E-SkewPolyF4}.
In~(1) we had seen that $x^2+1=(x+\omega)(x+\omega^2)$.
Right-multiplying the first factor by~$\omega^2$ and left-multiplying the second one by~$\omega$ does not change the product and thus
\[
     x^2+1=(x+\omega)(x+\omega^2)=(\omega x+1)(\omega x+1).
\]
Hence we have two factorizations of $x^2+1$ into linear polynomials.
While in the first factorization the two factors are relatively right-prime, the two factors of the second factorization are identical.
In the first factorization the linear factors are monic, whereas in the second one they are normalized
such that their constant coefficients are~$1$.
The reader is invited to check that Theorem~\ref{T-IrredFac} is indeed true: choose $\pi=\id$ and $h_1=\omega,\,h_2=\omega x$.
Examples like this will make it impossible to define multiplicities of roots for skew polynomials in a meaningful way.
We will comment on this at the end of Section~\ref{S-AlgSets}.
\end{exa}

We refer to \cite{Ore33} for further results on decompositions of skew polynomials, most notably completely reducible
polynomials, i.e.\
polynomials that arise as the least common left multiple of irreducible polynomials --- and thus generalize square-free commutative polynomials.

\section{Skew Polynomials and Linearized Polynomials}\label{S-Linearized}
In this short section we discuss the relation between skew-polynomial rings and the ring of linearized polynomials over finite fields.
The latter play an important role in the study of rank-metric codes.

Consider the ring $\Fxs$, where $\F=\F_{q^m}$ and~$\sigma$ is the $q$-Frobenius; see Example~\ref{E-FxsFinite}.
In the commutative polynomial ring $\F[y]$  define the subset
\[
  \cL:= \cL_{q^m,q}:=\Big\{\sum_{i=0}^N f_i y^{q^i}\,\Big|\, N\in\N_0, f_i\in \F\Big\}.
\]
Polynomials of this type are called $q$-\textbf{linearized}\index{$q$-linearized polynomials} because for any $f\in\cL$ the associated map $\F\longrightarrow\F,\,a\longmapsto f(a)$ is
$\F_q$-linear.
Linearized polynomials have been well studied in the literature and a nice overview of the basic properties can be found in \cite[Ch.~3.4]{LiNi97}.
In particular, $(\cL,+,\circ)$  is a (non-commutative) ring, where~$+$ is the usual addition and~$\circ$ the composition of polynomials.
The rings $\Fxs$ and $\cL$ are isomorphic in the obvious way.
Indeed, the map
\begin{equation}\label{e-Lambda}
  \Lambda: \Fxs\longrightarrow\cL,\quad \sum_{i=0}^N g_i x^i\longmapsto \sum_{i=0}^N g_i y^{q^i}
\end{equation}
is a ring isomorphism between $\Fxs$ and $(\cL,+,\circ)$.
The only interesting part of the proof is the multiplicativity of~$\Lambda$.
But that follows from
$\Lambda(ax^ibx^j)=\Lambda(a\sigma^i(b)x^{i+j})=\Lambda(ab^{q^i}x^{i+j})=ab^{q^i}y^{q^{i+j}}=ay^{q^i}\circ by^{q^j}$
for all $a,b\in\F$ and $i,j\in\N_0$.
As a consequence,~$\cL$ inherits all the properties of $\Fxs$ presented in the previous section.
We can go even further.
The polynomial $y^{q^m}-y\in\cL$ induces the zero map on~$\F=\F_{q^m}$.
Its pre-image under~$\Lambda$ is $x^m-1$, which is in the center of $\Fxs$; see Example~\ref{E-FxsFinite}.
As a consequence, the left ideal generated by $y^{q^m}-y$ is two-sided and gives rise to the quotient ring $\cL/(y^{q^m}-y)$.
Since the latter ring has cardinality $q^{m^2}$, this tells us that the quotient is isomorphic to the space of all $\F_q$-linear maps
on~$\F_{q^m}$. Thus,
\[
   \Fxs/(x^m-1)\cong \cL/(y^{q^m}-y)\cong\Mat_{m,m}(\F_q).
\]
Clearly, the second map is given by $g+(y^{q^m}-y)\longmapsto [g]_B^B$, where $[g]_B^B$ denotes the matrix representation of the map~$g$ with
respect to a chosen basis~$B$ of $\F_{q^m}$ over~$\F_q$.
Fix the basis $B=(b_0,\ldots,b_{m-1})$ and define its \textbf{Moore matrix}\index{Moore matrix} $S=\big(b_j^{q^i}\big)_{i,j=0}^{m-1}$
(called the \textbf{Wronskian}\index{Wronskian matrix} in \cite[(4.11)]{LaLe88}).
The linear independence of $b_0,\ldots,b_{m-1}$ implies that $S$ is in $\GL_m(\F_{q^m})$ (see \cite[Cor.~2.38]{LiNi97}).
Furthermore, in \cite[Lem.~4.1]{WuLi13} it is shown that for any $g=\sum_{i=0}^{m-1} g_iy^{q^i}$ we have $S\,[g]_B^B\,S^{-1}=D_g$, where
\begin{equation}\label{e-Dickson}
   D_g=\begin{pmatrix} g_0&g_1&\cdots &g_{m-2}&g_{m-1}\\ g_{m-1}^q&g_0^q&\cdots& g_{m-3}^q&g_{m-2}^q\\ \vdots&\vdots & &\vdots &\vdots\\
    g_1^{q^{m-1}}&g_2^{q^{m-1}}&\cdots& g_{m-1}^{q^{m-1}}&g_0^{q^{m-1}}\end{pmatrix}
\end{equation}
is the \textbf{Dickson matrix}\index{Dickson matrix} of~$g$.
This matrix is also known as \textbf{$q$-circulant}\index{$q$-circulant matrix} and generalizes the notion of a classical circulant matrix; see~\eqref{e-Circg}.
The isomorphisms above allow us to define the Dickson matrix of $g\in\Fxs$  as $D_g:=D_{\Lambda(g)}$
so that we obtain the ring isomorphism
\[
    \Fxs/(x^m-1)\longrightarrow \Mat_{m,m}(\F_q),\quad g+(x^m-1)\longmapsto D_g.
\]
Note that the $i^{\rm th}$ row of $D_g$ is given by the coefficient vector of $x^ig\in\Fxs$ reduced modulo $x^m-1$ via right division.
This will be made precise in the realm of skew circulants in Section~\ref{S-BSCC} (see Definition~\ref{D-CircMat} and the paragraph thereafter).

Linearized polynomials and their kernels play a crucial role in the study of the rank distance.
We refer to the vast literature on rank-metric codes, initiated by~\cite{Gab85}.
Most closely related to this survey are the articles \cite{CLU09,BoUl14,MP17,MP18} on the rank distance of skew-cyclic codes.

\section{Evaluation of Skew Polynomials and Roots}\label{S-Roots}
In this section we provide an overview of evaluating skew polynomials at field elements and roots of skew polynomials.
Ore in his seminal work~\cite{Ore33} did not define these concepts.
They were in fact introduced much later in 1986 by Lam, and the material of this and the next section is taken from the work of Lam and Leroy~\cite{Lam86,LaLe88,LaLe04,LLO08}.
Roots of skew polynomials as defined below in Definition~\ref{D-Evalf} also appeared already in the monograph~\cite[Sec.~8.5]{Co85} by Cohn in~1985.
(However, what is now commonly called a `right root' is considered a `left root' by Cohn.)

Throughout, we fix the skew-polynomial ring $\FFxs$.
Consider a polynomial $f=\sum_{i=0}^N f_i x^i\in \FFxs$.
Clearly, the usual notion of evaluating~$f$ at a point $a\in F$, that is, $f(a)=\sum_{i=0}^N f_ia^i$ is not well-defined if $\sigma\neq\text{id}$ because~$x$ does not commute with the field elements.
For instance, for the polynomial $f=bx=x\sigma^{-1}(b)$, where~$b$ is not fixed by~$\sigma$,
substituting a nonzero element~$a$ for~$x$ would lead to the contradiction  $ba=a\sigma^{-1}(b)$.

Circumventing this issue by requiring that coefficients be on the left when substituting~$a$ for~$x$ does not solve the problem
because it does not lead to a nice remainder theory.
Take for instance $f=x^3+\omega\in\F_4[x;\sigma]$, where $\F_4[x;\sigma]$ is as in Example~\ref{E-SkewPolyF4}.
Substituting $\omega$ for~$x$ yields $\omega^2$, and thus~$\omega$ is not a root in this naive sense.
Yet, one easily verifies that $x-\omega$ is a right divisor (and a left divisor) of~$f$.

A meaningful notion of evaluation of skew polynomials at field elements and roots of skew polynomials is obtained by making use of
division with remainder by the associated linear polynomials.
In the following we define right evaluation and right roots, and in this survey `root' will always mean `right root'.
The left-sided versions are analogous.

\begin{defi}\label{D-Evalf}
Let $f\in \FFxs$ and $a\in F$.
We define $f(a)=r$, where $r\in F$ is the remainder upon right division of~$f$ by $x-a$; that is,
$f=g\cdot(x-a)+r$ for some $g\in \FFxs$.
If $f(a)=0$, we call~$a$ a \textbf{(right) root}\index{Skew polynomial!Right root} of~$f$.
Thus $a$ is a root of~$f$ if and only if $(x-a)\rmid f$.
\end{defi}

Example~\ref{E-SkewPolyF4}(1) shows that a polynomial of degree~$N$ may have more than~$N$ roots.
If the field~$F$ is infinite, it may even have infinitely many roots.
For instance, in $\C[x;\sigma]$, where~$\sigma$ is complex conjugation, the polynomial $f=x^2-1$
splits as $(x+\ov{a})(x-a)$ for any $a$ on the unit circle, and thus has exactly the complex numbers on the unit circle
as roots.
The next example shows yet another surprising phenomenon, namely even if~$f$ splits into linear factors, it may only have one root.

\pagebreak

\begin{exa}\label{E-F8SoleRoot}
The polynomial $(x-\alpha^2)(x-\alpha)\in\F_8[x;\sigma]$, where $\alpha^3+\alpha+1=0$ and~$\sigma$ is the $2$-Frobenius, has the sole root~$\alpha$ in~$\F_8$.
Extending~$\sigma$ to the 2-Frobenius on $\F_{8^2}$, results in two additional roots of~$f\in\F_{8^2}[x;\sigma]$ in $\F_{8^2}\setminus\F_8$.
\end{exa}

The evaluation $f(a)$ can be computed explicitly without resorting to division with remainder.

\begin{defi}\label{D-Norm}
For any $i\in\N_0$ define  $N_i: F\longrightarrow F$ as $N_0(a)=1$ and $N_i(a)=\prod_{j=0}^{i-1}\sigma^j(a)$ for $i>0$.
We call $N_i$ the  \textbf{$i^{\rm th}$ norm on~$F$}.\index{$N_i$: $i^{\rm th}$ norm}
\end{defi}

Thus, $N_1(a)=a$ and $N_{i+1}(a)=N_{i}(a)\sigma^i(a)$  for all $a\in F$.
For $\F=\F_{q^m}$ and $\sigma$ the $q$-Frobenius, $N_m$ is simply the field norm of $\F$ over~$\F_q$.
Note that in the commutative case, i.e.,~$\sigma=\text{id}$, we simply have $N_i(a)=a^i$, and thus
the following result generalizes evaluation of commutative polynomials.

\begin{prop}[\mbox{\cite[Lem.~2.4]{LaLe88} or \cite[Eq.~(11) and Thm.~3]{Lam86}}]\label{P-faN}
Let  $f=\sum_{i=0}^N f_i x^i\in \FFxs$ and  $a\in F$.
Then
\[
    f(a)=\sum_{i=0}^N f_i N_i(a).
\]
\end{prop}

Now that we have a notion of roots for skew polynomials~$f\in\FFxs$ we may wonder about the relation to the roots of the associated
linearized polynomial $\Lambda(f)$ introduced in Section~\ref{S-Linearized}.
The latter are simply commutative polynomials and thus the ordinary notion of roots applies.

\begin{rem}\label{E-RootsLinear}
Consider $\FFxs=\Fxs$ as in Example~\ref{E-FxsFinite}, and let $g\in\Fxs$ and $\Lambda(g)\in\F[y]$ be as in~\eqref{e-Lambda}. Note that $\Lambda(g)(0)$ is always~$0$.
\begin{alphalist}
\item For any $q$ we have the following relation between the roots of~$g$ and~$\Lambda(g)$. For any $b\in \F^*$
        \[
             g(b^{q-1})=0\Longleftrightarrow \Lambda(g)(b)=0,
        \]
        see also \cite[Lem.~A.3]{GLNN18}.
        In order to see this, note that $\Lambda(g)(b)=0$ implies $\Lambda(g)(\alpha b)=0$ for all $\alpha\in\F_q$,
        which means that the linearized polynomial $y^q-b^{q-1}y$ is a divisor of~$\Lambda(g)$ in $(\F[y],+,\,\cdot\,)$.
        But then it is also a divisor of~$\Lambda(g)$ in the ring~$(\cL,+,\circ)$ (see \cite[Thm.~3.62]{LiNi97}), i.e., $\Lambda(g)=G\circ(y^q-b^{q-1}y)$ for some $G\in\cL$.
        Applying the ring homomorphism~$\Lambda^{-1}$ shows that $x-b^{q-1}$ is a right divisor of~$g$.
\item For $q=2$, Part~(a) shows that  the set of nonzero roots of~$\Lambda(g)$ coincides with the set of nonzero roots of~$g$.
\item If $q\neq2$, the roots of~$g$ do not agree with the roots of $\Lambda(g)$.
       To see this, take for instance $g=x-a$ for some nonzero $a\in\F_{q^m}$.
       Then~$a$ is a left and right root of~$g$.
       The associated linearized polynomial is $\Lambda(g)=y^q-ay=y(y^{q-1}-a)$, and it may or may not have nonzero roots.
       For instance, if $\F_{q^m}=\F_{3^2}$, the equation $y^2=a$ has two distinct roots for four values of~$a$ (the nonzero squares)
       and no roots for the other four values of~$a$.
       The discrepancy between roots of skew polynomials and roots of linearized polynomials is, of course, also related to the fact that multiplication in~$\cL$ is composition
       whereas a root~$c$ corresponds to a factor $y-c$ in the ordinary sense (which is not even a linearized polynomial).
\end{alphalist}
\end{rem}

There is an obvious relation between the right roots of a skew polynomial over a finite field
and the roots of a different associated commutative polynomial.
Consider again the skew-polynomial ring $\F[x;\sigma]$ as in Example~\ref{E-FxsFinite}.
Then the $i^{\rm th}$ norm is given by
\begin{equation}\label{e-NormFq}
           N_{i}(a)=a^{q^0+q^1+\cdots+q^{i-1}}=a^{\bracket{i}}, \ \text{ where }\bracket{i}:=\frac{q^i-1}{q-1}\text{ for }i\geq0.
\end{equation}
Proposition~\ref{P-faN} allows us to translate the evaluation of skew polynomials into evaluation of commutative polynomials (see also \cite[p.~278]{CLU09}).

\begin{rem}\label{R-CommPoly}
Define the map
\[
  \Fxs\longrightarrow \F[y],\quad \sum_{i=0}^n f_i x^i\longmapsto P_f:=\sum_{i=0}^n f_iy^{\bracket{i}}\in\F[y].
\]
Then $f(a)=P_f(a)$.
\end{rem}

Properties of the polynomial~$P_f$  can be found in \cite[Sec.~2]{Ler12}.
Unfortunately, the map $f\mapsto P_f$ does not behave well under multiplication and therefore the above result is of limited use.

Let us return to the general case with  an arbitrary field~$F$.
Having defined evaluation of polynomials, one may study properties of the map
\[
   \text{ev}_a: \FFxs\longrightarrow F,\quad f\longmapsto f(a).
\]
Clearly, this map is additive and left $F$-linear, but unlike in the commutative case, it is not multiplicative.
As an extreme case of this non-multiplicativity note that $f=x-a$ satisfies $f(a)=0$, whereas $(fb)(a)\neq0$ for any $b\in\F$ not fixed by~$\sigma$.
Yet, evaluation is close to being multiplicative.
Before we can make this precise we need the following definition.

\begin{defi}\label{D-Conj}
Let $a\in F$.  For $c\in F^*$ we define $a^c:=\sigma(c)ac^{-1}$.
We say that $a,b\in F$ are $\sigma$-\textbf{conjugate}\index{$\sigma$-conjugate}\index{Conjugate}
 if $b=a^c$ for some $c\in F^*$.
The $\sigma$-\textbf{conjugacy class} of~$a$ is
\index{$\sigma$-conjugacy class}\index{Conjugacy class}\index{$\Delta(a)$: Conjugacy class of~$a$}
\[
   \Delta(a)=\{a^c\mid c\in F^*\}.
\]
\end{defi}

Since the automorphism is fixed throughout this text, we will drop the prefix~$\sigma$.
The reader should be aware of the ambiguity of the notation.
For instance, for $c=-1\in F$, the notation $a^c$ does not represent the inverse of~$a$;
in fact, in this case $a^c=a$.
More generally, $a^{-c}=a^c$ for any $c\in F^*$.

It is easy to see that conjugacy defines an equivalence relation.
Moreover, $\Delta(0)=\{0\}$, and if $\sigma=\id$, then $\Delta(a)=\{a\}$ for all $a\in F$.
Furthermore, if $a\neq0$, then $a^c=a$ iff $c$ is in the fixed field of~$\sigma$.
The relevance of conjugacy for us stems from the equivalence
\[
  b=a^c\Longleftrightarrow (x-b)c=\sigma(c)(x-a)
\]
and the identities~\cite[(3.3)]{LaLe04}
\begin{equation}\label{e-lclmLinear}
  \lclm(x-a,x-b)=(x-b^{b-a})(x-a)=(x-a^{a-b})(x-b)\text{ for any }a\neq b,
\end{equation}
which tell us that, up to conjugacy, linear factors can be reordered.
Furthermore, as a special case of Proposition~\ref{P-CfCg} we have
\[
   \text{$x-a$ and $x-b$ are similar in the sense of Definition~\ref{D-similar} }\Longleftrightarrow a,\,b\text{ are conjugate}.
\]

\begin{exa}\label{E-ConjClasses}
\begin{alphalist}
\item For any finite field $F=\F_{q^m}$ with $q$-Frobenius~$\sigma$, the identity $a^c=c^{q-1}a$ implies that
the nonzero conjugacy classes are given by the cosets of $\Delta(1)=\{c^{q-1}\mid c\in F^*\}$ in $F^*$.
Thus, for $q=2$ the conjugacy classes are $\{0\}$ and $\Delta(1)=F^*$, whereas for $q=3$ there are two nonzero
conjugacy classes, one of which consists of the squares of~$F^*$ and the other of the non-squares.
\item For~$\C$ with complex conjugation, the nonzero conjugacy classes are exactly the circles about the origin.
\end{alphalist}
\end{exa}

\begin{rem}\label{R-SizeConj}
Obviously, the conjugacy classes are the orbits of the group action $F^*\times F\longrightarrow F,\ (c,a)\longmapsto a^c$,
and the stabilizer of any nonzero~$a$ is the multiplicative group of the fixed field of~$\sigma$.
Therefore, in the case where $F=\F_{q^m}$ and $\sigma$ is the $q$-Frobenius,
the nonzero conjugacy classes have size $(q^m-1)/(q-1)$.
This also shows that there are $q$ conjugacy classes (including $\{0\}$).
\end{rem}

Now we can formulate the product theorem.
It appeared first in \cite[Thm.~2]{Lam86} and was later extended to more general skew-polynomial rings in \cite[Thm.~2.7]{LaLe88}.
The proof follows by direct computations using Proposition~\ref{P-faN} and properties of the norms $N_i$.

\begin{theo}\label{T-ProdEval}
Let $f,g\in \FFxs$ and $a\in F$. Then
\[
   (fg)(a)=\left\{\begin{array}{cl}0,&\text{if }g(a)=0\\ f(a^{g(a)})g(a),&\text{if }g(a)\neq0.\end{array}\right.
\]
In particular, if $a$ is root of $fg$, but not of~$g$, then the conjugate~$a^{g(a)}$ is a root of~$f$.
\end{theo}

We have seen already that the number of roots of a skew polynomial may vastly exceed its degree.
Taking conjugacy into account, however, provides us with the following generalization of the commutative case.
It follows quickly from the previous result by inducting on the degree.

\begin{theo}[\mbox{\cite[Thm.~4]{Lam86}}]\label{T-RootsConjClasses}
Let $f\in \FFxs$ have degree~$N$.
Then the roots of~$f$ lie in at most~$N$ distinct conjugacy classes.
Furthermore, if $f=(x-a_1)\cdots(x-a_N)$ for some $a_i\in F$ and $f(a)=0$, then~$a$ is conjugate to some~$a_i$.
\end{theo}

Note that for $F=\F_{q^m}$ with $q=2$, the theorem does not provide any insight because in this case there is only one
conjugacy class. This also shows that the converse of Theorem~\ref{T-RootsConjClasses} is not true:
not every conjugate of some~$a_i$ is a root of~$f$ (take $N=1$ for instance).

Even more,  the above theorem does not state that every conjugacy class $\Delta(a_i)$
contains a root of~$f$. This is indeed in general not the case, and an example can be found in the skew-polynomial ring
$\Q(t)[x;\sigma]$, where $\sigma$ is the $\Q$-algebra automorphism given by $t\mapsto t+1$.
However, for finite fields the last statement is in fact true.

\begin{theo}\label{T-RootsConjClasses2}
Consider the ring $\F[x;\sigma]$ as in Example~\ref{E-FxsFinite}, and let $f=(x-a_1)\dotsm(x-a_N)$ for some $a_i\in\F^*$.
Then each conjugacy class $\Delta(a_i)$ contains a root of~$f$.
\end{theo}

\noindent\textbf{Proof:}
It suffices to show that $\Delta(a_1)$ contains a root of~$f$, which means that~$f$ is of the form
$f=(x-b_1)\dotsm(x-b_{N-1})(x-a_1^c)$ for some $b_1,\ldots,b_{N-1},c\in\F^*$.
Since $(a^{c_1})^{c_2}=a^{c_1c_2}$, the case $N=2$ is sufficient.
Thus, let $f=(x-b)(x-a)$ for some $a,b\in\F^*$.
If $a\in\Delta(b)$ there is nothing to prove. Thus let $a\not\in\Delta(b)$.
Let now $c\in\F^*$.
Theorem~\ref{T-ProdEval} implies $f(b^c)=((b^c)^{b^c-a}-b)(b^c-a)$, and thus $f(b^c)=0$ iff $(b^c)^{b^c-a}=b$.
It is easy to see that the latter is equivalent to $\sigma(\sigma(c)b-ac)=\sigma(c)b-ac$, which in turn is equivalent to
$\sigma(c)b-ac\in\F_q$.
Thus we have to establish the existence of some $c\in\F^*$ such that $\sigma(c)b-ac\in\F_q$.
Consider the $\F_q$-linear map $\Psi_{a,b}:\F\longrightarrow\F,\  c\mapsto \sigma(c)b-ac$.
If we can show that $\Psi_{a,b}$ is injective, then it is bijective and we are done.
Suppose there exists $d\in\ker\Psi_{a,b}\setminus\{0\}$. Then $\sigma(d)b-ad=d^qb-ad=0$, hence $d^{q-1}=a/b$.
But then $1=d^{q^m-1}=(d^{q-1})^{\bracket{m}}=(a/b)^{\bracket{m}}$.
This shows that $\Psi_{a,b}$ is injective if $(a/b)^{\bracket{m}}\neq1$.
On the other hand, if $(a/b)^{\bracket{m}}=1$, then the order of $a/b$ in~$\F^*$, say~$t$, is a divisor of $\bracket{m}$.
Furthermore, $a/b=\omega^{k(q^m-1)/t}$ for some~$k$ and a primitive element~$\omega$.
Writing $ts=\bracket{m}$, we conclude $a/b=(\omega^{ks})^{q-1}$.
But this means that $a$ and~$b$ are conjugate (see Example~\ref{E-ConjClasses}(a)), a contradiction.
\hfill$\Box$

A similar reasoning shows that the previous result is also true in the skew-polynomial ring $\C[x;\sigma]$ with complex conjugation~$\sigma$.

\section{Algebraic Sets and Wedderburn Polynomials}\label{S-AlgSets}
We now further the theory of right roots of skew polynomials by introducing minimal polynomials and algebraic sets and presenting some of their properties.
The material is  again from~\cite{Lam86,LaLe04}.
Throughout, we fix a skew-polynomial ring $\FFxs$.

\begin{defi}\label{D-AlgSet}
For a polynomial $f\in \FFxs$ denote by $V(f)$ its set of (right) roots in~$F$; thus $V(f)=\{a\in F\mid f(a)=0\}$.
We call $V(f)$ the \textbf{vanishing set} of~$f$.\index{Vanishing set $V(f)$}
A subset $A\subseteq F$ is called $\sigma$-\textbf{algebraic}
if there exists some nonzero $f\in \FFxs$ such that
$A\subseteq V(f)$; that is, $f$ vanishes on~$A$.\index{$\sigma$-algebraic set}\index{Algebraic set}
In this case, the monic polynomial of smallest degree, say~$f$, such that  $A\subseteq V(f)$, is uniquely determined by~$A$ and called the
$\sigma$-\textbf{minimal polynomial}\index{Minimal polynomial of a set} of~$A$, denoted by $m_A$.\index{$m_A$: Minimal polynomial of set~$A$}
The degree of~$m_A$ is called the $\sigma$-\textbf{rank} of~$A$, denoted by $\rank(A)$.\index{$\rank(A)$: Rank of a set}
\end{defi}

Again, we will drop the prefix~$\sigma$ as there will be no ambiguity.
The well-definedness of the minimal polynomial is a consequence of Theorem~\ref{T-ProdEval}  along with the fact that  $\FFxs$
is a  left principal ideal ring.

The reader may wonder why an algebraic set has to be merely contained in a vanishing set, but not necessarily be
a vanishing set itself.
The latter would be too restrictive for a meaningful theory because in general the minimal polynomial~$m_A$ of a set~$A$ has additional roots outside~$A$ (see Example~\ref{E-ConjV}(b)).
This phenomenon gives rise to the  \textbf{closure} of~$A$, defined as the vanishing set $V(m_A)$.
\index{Closure of a set}
However, we do not need this notion and therefore will not discuss it in further detail.

We begin with discussing the vanishing sets $V(f)$ for given polynomials.
First of all,
\[
  V(f)\subseteq V(g)\Longrightarrow V(fh)\subseteq V(gh)
\]
for all $f,g,h\in\FFxs$.
On first sight the implication may feel counterintuitive because $V(f)$ denotes the set of right roots.
Yet the reader can readily verify that it is just a simple consequence of Theorem~\ref{T-ProdEval}.
On the other hand, the analogous statement with left factors~$h$ is not true in general; that is, there exists $f,g,h\in\FFxs$ such that
\[
  V(f)\subseteq V(g)\ \text{ and }\ V(hf)\not\subseteq V(hg).
\]

\begin{exa}\label{E-Vhf}
Consider $\F_4[x;\sigma]$ from Example~\ref{E-SkewPolyF4}.
Let $f=h=x+1$ and $g=x^2+\omega^2x+\omega$.
Then one easily checks that $V(f)=\{1\}=V(g)$.
In  Example~\ref{E-SkewPolyF4}(1) we have seen that $V(hf)=\{1,\omega,\omega^2\}$.
However, $hg=x^3+\omega^2x^2+\omega$ has sole root~$1$.
\end{exa}

We now turn to algebraic sets.
Obviously, in the commutative case, i.e.\ $\sigma=\id$,  algebraic sets are exactly the finite sets.
This is not the case for skew polynomials as we have seen right after Definition~\ref{D-Evalf} for $\C[x;\sigma]$ with complex conjugation~$\sigma$.
From~\eqref{e-lclmLinear} we deduce that every set $A=\{a,b\}$ of cardinality~$2$ has rank~$2$, whereas
Example~\ref{E-SkewPolyF4}(1) provides us with a set of cardinality~$3$ and rank~$2$.
In general, a finite set $A=\{a_1,\ldots,a_n\}$ has minimal polynomial $m_A=\lclm(x-a_1,\ldots,x-a_n)$.
Using induction on the cardinality and the degree formula in Proposition~\ref{R-PropR}(d), one obtains immediately:

\begin{prop}[see also \mbox{\cite[Prop.~6]{Lam86}}]\label{P-FiniteAlg}
Let $A=\{a_1,\ldots,a_n\}\subseteq F$. Then $A$ is algebraic and $\rank(A)=:r\leq |A|$.
Furthermore, there exist distinct $b_1,\ldots,b_r\in A$ such that
$m_A=\lclm(x-b_1,\ldots,x-b_r)$.
\end{prop}

\begin{exa}[\mbox{\cite[Rem.~2.4]{Ler12}}]\label{E-MinPolyF}
Let $A=\F_{p^r}$, where~$p$ is prime and $r\in\N$.
Then $m_A=\lclm(x-a\mid a\in\F_{p^r})=x^{r(p-1)+1}-x$. Thus, $\rank(\F_{p^r})=r(p-1)+1$.
\end{exa}

Here is a particularly interesting example.
The proof is analogous to the commutative case.

\begin{exa}[\mbox{\cite[Prop.~4]{BoUl14}}]\label{E-MinPolya}
Let $\F_{q^s}$ be an extension field of $\F=\F_{q^m}$ and consider $\F_{q^s}[x;\sigma]$ with $q$-Frobenius~$\sigma$.
Fix an element $a\in\F_{q^s}$ and set $A=\{\tau(a)\mid \tau\in\Aut(\F_{q^s}\mid\F_{q^m})\}$.
Then the minimal polynomial~$m_A$ is in $\Fxs$ and is the nonzero monic polynomial of smallest degree in $\Fxs$ with (right) root~$a$.
It is called the $\sigma$-\textbf{minimal polynomial} of~$a$ over~$\F$.
\end{exa}

The minimal polynomial of an algebraic set always factors linearly. The following result comes closest to the commutative case.

\begin{prop}[\mbox{\cite[Lem.~5]{Lam86}}]\label{P-AlgMiPo}
Let $A\subseteq F$ be an algebraic set of rank~$r$. Then its minimal polynomial is of the form
$m_A=(x-a_1)\cdots(x-a_r)$, where each~$a_i$ is conjugate to some $a\in A$.
\end{prop}

This result is indeed the best possible in the sense that the roots of the linear factors need not be in~$A$.

\begin{exa}\label{E-ConjV}
\begin{alphalist}
\item Consider $\F_{3^3}[x;\sigma]$ with $3$-Frobenius~$\sigma$ and primitive element $\beta$ satisfying $\beta^3+2\beta+1=0$.
        Let $A=\{\beta^{14},\beta^{25}\}$.
        Then $m_A=x^2+\beta x+\beta=(x-\beta^{13})(x-\beta^{14})=(x-\beta^2)(x-\beta^{25})$, and thus~$m_A$ is not the product
        of two linear terms with roots in~$A$.
        With the aid of Example~\ref{E-ConjClasses} one concludes that $\beta^{13}$ is conjugate to $\beta^{25}$ and $\beta^2$ is
        conjugate to~$\beta^{14}$. Finally, $A=V(m_A)$; that is,~$m_A$ has no further roots in~$\F_{3^3}$.
\item The linear factors $x-a_i$ in Proposition~\ref{P-AlgMiPo} need not be distinct.
        Consider for instance $\F_{2^4}[x;\sigma]$ with $2$-Frobenius~$\sigma$ and primitive element~$\gamma$ satisfying $\gamma^4+\gamma+1=0$.
        The polynomial
        \[
            f=(x-\gamma^2)(x-\gamma^{12})(x-\gamma^2)=(x-\gamma^3)(x-\gamma^{14})(x-\gamma^{14})=
            x^3+\gamma^7x^2+\gamma^3x+\gamma
         \]
         is the minimal polynomial of
        $A=\{1,\gamma^2,\gamma^3,\gamma^6,\gamma^8,\gamma^{13},\gamma^{14}\}$ and in fact $V(f)=A$.
        In order to illustrate Proposition~\ref{P-FiniteAlg} we mention that $f=\lclm(x-1,x-\gamma^2,x-\gamma^3)$,  which
        shows that the set $B=\{1,\gamma^2,\gamma^3\}$ is algebraic, but not a vanishing set itself:
        every polynomial vanishing on~$B$ has additional roots in~$\F_{2^4}$.
        On the other hand, $f\neq\lclm(x-1,x-\gamma^2,x-\gamma^8)$. The latter polynomial is given by
        $g=x^2+\gamma^5x+\gamma^{10}$.
       \end{alphalist}
\end{exa}

The rank of a finite algebraic set can be determined via the `skew version' of the classical Vandermonde matrix.
The \textbf{skew Vandermonde matrix} has been introduced by
Lam in \cite[p.~194]{Lam86}.
For $a_1,\ldots,a_r\in F$ it is the matrix in ${\rm Mat}_{n,r}(F)$ defined as
\index{Skew Vandermonde matrix} \index{$V_n(a_1,\ldots,a_r)$: Skew Vandermonde matrix}
\begin{equation}\label{e-Vand}
  V_n^\sigma(a_1,\ldots,a_r):=V_n(a_1,\ldots,a_r)=\begin{pmatrix}1&\cdots&1\\ N_1(a_1)&\cdots&N_1(a_r)\\ \vdots& &\vdots\\ N_{n-1}(a_1)&\cdots&N_{n-1}(a_r)\end{pmatrix}.
\end{equation}
The skew Vandermonde matrix depends on~$\sigma$ (because the norms do).
Proposition~\ref{P-faN} shows that for $g=\sum_{i=0}^{n-1}g_i x^i\in F[x;\sigma]$ we have
\[
    (g(a_1),\ldots,g(a_r))=(g_0,\ldots,g_{n-1})V_n^\sigma(a_1,\ldots,a_r).
\]
Using~\eqref{e-NormFq}, we conclude that for the skew-polynomial ring $\Fxs$, where $\sigma$ is the $q$-Frobenius, the skew Vandermonde matrix evaluates
the powers $x^{\bracket{0}},x^{\bracket{1}},\ldots,x^{\bracket{n-1}}$ at $a_1,\ldots,a_r$.
This matrix must not be confused with the Moore matrix of $a_1,\ldots,a_r$, which evaluates the powers $x^{q^0},x^{q^1},\ldots,x^{q^{n-1}}$ at
$a_1,\ldots,a_r$.
The relation between the Moore matrix and an associated Vandermonde  matrix is spelled out in Example~\ref{E-NormBasisVand} below.
The following results are not hard to show (in~\cite{Lam86} a bit more work is required because the coefficients are from a division ring).

\begin{theo}[\mbox{\cite[Thm.~8]{Lam86}}]\label{T-VandRank}
Let $A=\{a_1,\ldots,a_n\}\subseteq F$. Then $\rank(A)=\rank(V_n(a_1,\ldots,a_n))$.
As a consequence, if $\rank(A)=|A|$ (such a set is called P-independent),  then $\rank(B)=|B|$ for every subset $B\subseteq A$.
\end{theo}

\begin{exa}\label{E-VandCompl}
\begin{alphalist}
\item Consider $\C[x;\sigma]$ with complex conjugation~$\sigma$.
Let $A=\{a_1,\ldots,a_n\}\subseteq\C$, where $a_1,\ldots,a_n$ are not all equal and $|a_1|=\cdots=|a_n|=:c$.
Then $N_2(a_i)=a_i\sigma(a_i)=c^2$ for all~$i$.
This implies that $V_n(a_1,\ldots,a_n)$ has rank~$2$, and this is consistent with the fact that $m_A=x^2-c^2=(x+a_i)(x-\ov{a_i})$.
\item Consider Example~\ref{E-ConjV}(b). Then
        \[
            V_3(1,\gamma^2,\gamma^8)=\begin{pmatrix}1&1&1\\ 1&\gamma^2&\gamma^8\\1&\gamma^6 &\gamma^9 \end{pmatrix}
        \]
        has rank~$2$, consistent with the fact that $\lclm(x-1,x-\gamma^2,x-\gamma^8)$ has degree~$2$.
\end{alphalist}
\end{exa}

\begin{exa}\label{E-NormBasisVand}
Consider $\Fxs$ as in Example~\ref{E-FxsFinite}, thus $\F=\F_{q^m}$.
Suppose $a_0,\ldots,a_{m-1}$ is a basis of~$\F_{q^m}$ over~$\F_q$. 
Then one easily verifies (\cite[Eq.~(4)]{BoUl14} or \cite[Thm.~5]{LMK15}) that
\[
  V_m(a_0^{q-1},\ldots,a_{m-1}^{q-1})\begin{pmatrix}a_0& & & \\ &\!\!a_1& & &\\ & &\!\!\ddots\!\! & \\ & & &a_{m-1}\!\!\end{pmatrix}
  =\begin{pmatrix}a_0&\cdots&a_{m-1}\\ a_0^q&\cdots&a_{m-1}^q\\ \vdots & &\vdots\\ a_0^{q^{m-1}}&\cdots&a_{m-1}^{q^{m-1}}\end{pmatrix}.
\]
Note that the matrix on the right hand side is the Moore matrix of $a_0,\ldots,a_{m-1}$ (see Section~\ref{S-Linearized}).
Since the latter is invertible thanks to the linear independence of $a_0,\ldots,a_{m-1}$ (\cite[Cor.~2.38]{LiNi97}), the same is true for the Vandermonde matrix on the left hand side.
Let us now apply this to a normal basis $\{\gamma,\gamma^q,\ldots,\gamma^{q^{m-1}}\}$ of $\F_{q^m}$.
Then the Vandermonde matrix above is $V_m(b,b^q,\ldots,b^{q^{m-1}}\big)$, where $b=\gamma^{q-1}$.
 It is easy to see that $N_i(b^{q^j})=(\gamma^{-1})^{q^j}\gamma^{q^{i+j}}$, which in turn yields that
$b^{q^j}$ is a right root of $x^m-1\in\Fxs$ for all $j=0,\ldots,m-1$.
As a consequence, $x^m-1=\lclm(x-b,x-b^q,\ldots,x-b^{q^{m-1}})$ thanks to Theorem~\ref{T-VandRank}.
Note that the obvious right root~$1$ of $x^m-1$ does not appear in the list $b,b^q,\ldots,b^{q^{m-1}}$.
Theorem~\ref{T-VandRank} also shows for any subset $\{j_1,\ldots,j_r\}\subseteq\{0,\ldots,m-1\}$ the polynomial
$\lclm(x-b^{q^{j_1}},\ldots,x-b^{q^{j_r}})$ has degree~$r$ (see also \cite[Lem.~3.1]{GLN18}).
Finally, consider a polynomial $x^m-a\in\Fxs$ for some $a\in\F^*$ and suppose $c\in\F$ is a root of $x^m-a$.
Using the multiplicity of the maps $N_i$, one easily deduces that $x^m-a= \lclm(x-cb,x-cb^q,\ldots,x-cb^{q^{m-1}})$.
\end{exa}

We now turn to polynomials that occur as minimal polynomials of an algebraic set~$A$.

\begin{defi}\label{D-Wedd}
A monic polynomial $f\in\FFxs$ is a \textbf{Wedderburn polynomial over~$F$}\index{Wedderburn polynomial}
or simply \textbf{W-polynomial} \index{W-polynomial} if
$f=m_A$ for some $A\subseteq F$.
\end{defi}

The polynomial $x^2+1$ is a W-polynomial over~$\F_4$ as it is $m_{\{1,\omega,\omega^2\}}$ (see Example~\ref{E-SkewPolyF4}(1)), but it is not a W-polynomial
over~$\F_2$. Thus the field~$F$ matters in the definition of a W-polynomial.
We will always assume that the field is the coefficient field of the skew polynomial ring under consideration.
In general, $x^m-1\in\F_{q^m}[x;\sigma]$, where~$\sigma$ is the $q$-Frobenius, is a W-polynomial by Example~\ref{E-NormBasisVand}.

The polynomial $x^2-1\in\C[x;\sigma]$, where~$\sigma$ is complex conjugation, is the minimal polynomial of the unit circle (or of the set $\{1,-1\}$) and thus
a W-polynomial.
In $\F_4[x;\sigma]$ the polynomial $f=(x+1)(x+\omega)$ is not a W-polynomial because $V(f)=\{\omega\}$.

If the algebraic set~$A$ is finite, say $A=\{a_1,\ldots,a_N\}$, then Theorem~\ref{T-ProdEval} implies that $f=m_A=\lclm(x-a_1,\ldots,x-a_N)$.
This shows that in the commutative case, W-polynomials are simply the separable polynomials that factor linearly.
One may also note that W-polynomials are a special case of completely reducible polynomials in the sense of \cite[p.~495]{Ore33} by Ore.
The latter are defined as the least common left multiple of irreducible polynomials.

Since $m_{V(f)}\rmid f$ for any $f\in\FFxs$, we observe that $f$ is a W-polynomial iff $m_{V(f)}=f$.
If $\deg(f)=N$, then this reads as ``$f$ is Wedderburn iff $\rank(V(f))=N$'', which may be understood as~$f$ having sufficiently many roots (see \cite[Prop.~3.4]{LaLe04}).

Let us list some properties of W-polynomials.

\pagebreak

\begin{prop}[\mbox{\cite[Prop.~4.3]{LaLe04}}]\label{R-Wedd}
Let $A,B\subseteq F$ be algebraic sets.
\begin{alphalist}
\item $m_{A\cup B}=\lclm(m_A,m_B)$; thus $\rank(A\cup B)\leq \rank(A)+\rank(B)$.
\item $\rank(A\cup B)=\rank(A)+\rank(B)\Longleftrightarrow\gcrd(m_A,m_B)=1\Longleftrightarrow V(m_A)\cap V(m_B)=\emptyset$.
\end{alphalist}
\end{prop}

Part~(a) as well as the first equivalence in~(b) are clear; see also Theorem~\ref{R-PropR}(d). The second part in~(b) requires more work.
We now present some strong properties of $W$-polynomials.
More machinery is needed to derive most of them, especially the $\Phi$-transform and $\lambda$-transform introduced by Lam/Leroy in~\cite{LaLe04}, and we refer
to the excellent presentation in \cite[Sec.~4 and 5]{LaLe04} for further details.

\begin{theo}[\mbox{\cite[Thms.~5.1, 5.3, 5.9, 5.10]{LaLe04}}]\label{T-WPoly}\
\begin{arabiclist}
\item Let $f\in\FFxs$ be a monic polynomial of degree~$N$. The following are equivalent.
        \begin{romanlist}
        \item $f$ is a W-polynomial.
        \item $f=\lclm(x-a_1,\ldots,x-a_N)$ for some distinct elements $a_i\in F$.
        \item $f$ splits completely and every monic factor of~$f$ is a W-polynomial.
        \end{romanlist}
\item Let $f,\,g\in\FFxs$ be similar monic polynomials and~$f$ a W-polynomial. Then $g$ is a W-polynomial.
\item Let $g,h$ be W-polynomials. The following are equivalent.
        \begin{romanlist}
        \item $gh$ is a W-polynomial.
        \item $1\in\lideal{g}+\rideal{h}$.
        \item $\{k\in\FFxs\mid gk\in\lideal{g}\}\subseteq\lideal{g}+\rideal{h}$.
        \end{romanlist}
\end{arabiclist}
\end{theo}

The set on the left hand side of (3)(iii) above is called the idealizer of $\lideal{g}$. It is the largest subring of $\FFxs$ in which $\lideal{g}$ is a two-sided ideal.

We close this section with a brief discussion of `multiplicities of roots' and `splitting fields' for skew polynomials.

It is tempting to define the multiplicity of a root~$a$ of the skew polynomial $f\in\FFxs$ as the largest exponent~$r$ for which $(x-a)^r$ is a right divisor of~$f$.
However, this defines the multiplicity based on the \emph{monic} linear factor with root~$a$.
Unfortunately, rescaling a linear factor with root~$a$ from the left (or from the right) may change the exponent for the right divisor of~$f$.
Indeed, in Example~\ref{E-x21} we have seen that $x+\omega^2$ appears with exponent~$1$ as right divisor of $x^2+1$, whereas
$\omega(x+\omega^2)=\omega x+1$ appears with exponent~$2$.
For this reason it is not meaningful to define the multiplicity of roots in this way.
The reader may also note that over a finite field~$\F$ the commutative polynomial~$P_f$ from Remark~\ref{R-CommPoly} has only simple roots (if $f_0\neq0$)
and thus cannot serve for the definition of multiplicity either.
As to our knowledge no notion of multiplicity of roots for skew polynomials has been discussed in the literature.

It should not come as a surprise that also the notion of a splitting field is questionable.
First of all, when considering extension fields of~$F$ we also need to extend the automorphism~$\sigma$.
Of course, this is not unique in general.
But even if we extend the $q$-Frobenius of~$\F_{q^m}$ to the $q$-Frobenius on an extension field $\F_{q^M}$, we still may ask whether
we want a splitting field to be a `smallest' extension in which the given polynomial splits or whether we want it a `smallest' one in which the polynomial has `all its roots'?
Example~\ref{E-F8SoleRoot} has already shown that these two objectives are not identical.
The latter objective is easy to achieve for the finite field case $\F_{q^m}[x;\sigma]$: from Remark~\ref{R-CommPoly} it is clear that the splitting field of the
commutative polynomial $P_f\in\F_{q^m}[y]$ is the smallest field that contains all the roots of~$f$.
On the other hand, the polynomial $x^2-i\in\C[x;\sigma]$ with complex conjugation~$\sigma$ has no roots in~$\C$ (since
$N_2(c)=|c|^2$ is real for any $c\in\C$), and thus its splitting field, if it exists, must be a  (transcendental) field extension of~$\C$.
As to our knowledge no theory of splitting fields has been developed for skew polynomials.

\section{A Circulant Approach Toward Cyclic Block Codes}\label{S-CBC}
We briefly summarize the algebraic theory of classical cyclic block codes with the aid of circulant matrices.
This is standard material of any introductory course on block codes (see also \cite[Ch.~4]{HP03}),
and is presented here for the mere purpose to serve as a guideline for the skew case in the next sections.

Let us fix a finite field~$\F=\F_q$ and a length~$n$. Usually, one requires that~$q$ and~$n$ be relatively prime, but that is not needed for the algebraic theory.
It only plays a role when it comes to distance considerations.
We consider the quotient ring $\cR=\F[x]/(x^n-1)$, which as an $\F$-vector space is isomorphic to~$\F^n$ via the map
\[
  \pp: \F^n\longrightarrow \cR,\quad (g_0,\ldots, g_{n-1})\longmapsto \overline{\sum_{i=0}^{n-1}g_ix^i},
\]
where $\ov{\,\cdot\,}$ denotes cosets.
Set $\vv:=\pp^{-1}$ (and think of these maps as \emph{vectorization} and \emph{polynomialization}).

A \textbf{cyclic code} in $\F^n$ is, by definition, a subspace of the form $\cC=\vv(I)$, where~$I$ is an ideal in~$\cR$.
This means that~$\cC$ is invariant under the cyclic shift $(a_0,\ldots, a_{n-1})\mapsto (a_{n-1},a_0,\ldots, a_{n-2})$.
Usually, the code is identified with its corresponding ideal so that its codewords are polynomial of degree less than~$n$.
It follows from basic algebra that~$\cR$ is a principal ideal ring and every ideal is of the form
$(\ov{g})$, where~$g$ is a monic divisor of~$x^n-1$.
Such a generator is unique and called the \textbf{generator polynomial} of the cyclic code $(\ov{g})$.
We conclude that the number of cyclic codes of length~$n$ over~$\F$ equals the number of divisors of $x^n-1$.
If $gh=x^n-1$, then~$h$ is the \textbf{parity check polynomial} of~$\cC$.
For all of this see also \cite[Thm.~4.2.1,~4.2.7]{HP03}.

Before summarizing some well-known facts about cyclic codes we introduce circulant matrices.
For any $g=(g_0,\ldots, g_{n-1})\in\F^n$ define the \textbf{circulant}\index{Circulant}
\begin{equation}\label{e-Circg}
\Gamma(g):=\begin{pmatrix}g_0&g_1&\ldots&g_{n-2}&g_{n-1}\\
                       g_{n-1}&g_0&\ldots&g_{n-3}&g_{n-2}\\
                        \vdots&\vdots&  &\vdots &\vdots\\
                        g_2&g_3&\ldots&g_0&g_1\\
                        g_1&g_2&\ldots&g_{n-1}&g_0
          \end{pmatrix}.
\end{equation}
Indexing by $i=0,\ldots,n-1$, we see that the $i^{\rm th}$ row of $\Gamma(g)$ is given by $\vv(\ov{x^i\sum_{j=0}^{n-1}g_jx^j})$.
The set $\CircM:=\{\Gamma(g)\mid g\in\F^n\}$ is an $n$-dimensional subspace of~$\Mat_{n,n}(\F)$.
We also define
\begin{equation}\label{e-Recip}
  \rho: \F[x]\longrightarrow\F[x],\quad g=\sum_{i=0}^r g_ix^i\longmapsto x^rg(x^{-1})=\sum_{i=0}^r g_i x^{r-i}\ \text{ (where $g_r\neq0$)}.
\end{equation}
The image $\rho(g)$ is called the \textbf{reciprocal} of~$g$.\index{Reciprocal of a polynomial}

The following properties of circulant matrices are either trivial or
well-known; see~\cite[Thm.~4]{KrSi12}, ~\cite[p.~501]{MS77} or~\cite{Da79} for a general reference on
circulant matrix theory.

\begin{rem}\label{R-MgMh}
\begin{alphalist}
\item The map ${\rm Circ}\longrightarrow\cR,\ \Gamma(g_0,\ldots, g_{n-1})\longmapsto \ov{\sum_{i=0}^{n-1}g_ix^i}$,
      is an $\F$-algebra isomorphism.
      We may and will use the notation $\Gamma(\ov{\sum_{i=0}^{n-1}g_ix^i})$ for $\Gamma(g_0,\ldots, g_{n-1})$.
      Then the above tells us, among other things, that $\Gamma(\ov{g}\ov{h})=\Gamma(\ov{g})\Gamma(\ov{h})=\Gamma(\ov{h})\Gamma(\ov{g})$.
\item $\rank\, \Gamma(\ov{g})=\deg\frac{x^n-1}{\gcd(g,x^n-1)}=:k$
      (where the quotient is evaluated in $\F[x]$) and
      every set of $k$ consecutive rows (resp.\ columns) of $\Gamma(\ov{g})$ is linearly
      independent.
\item The map $\phi: \cR\longrightarrow\cR,\ \ov{g}\longmapsto\ov{g(x^{n-1})}$ is a well-defined involutive $\F$-algebra automorphism corresponding to transposition in~$\CircM$, i.e.,
       $\Gamma(\ov{g})\T=\Gamma(\phi(\ov{g}))$.
\item Let $g=\sum_{i=0}^r g_i x^i$ be of degree~$r$.
        Then $\ov{x^r}\phi(\ov{g})=\ov{\rho(g)}$ or, equivalently, $\phi(\ov{g})=\ov{x^{n-r}\rho(g)}$.
        Hence $\Gamma(\ov{\rho(g)})=\Gamma(\ov{x^r})\Gamma(\phi(\ov{g}))=\Gamma(\phi(\ov{g}))\Gamma(\ov{x^r})$ and
        since $\ov{x^r}$ is a unit, the circulants $\Gamma(\phi(\ov{g}))$ and $\Gamma(\ov{\rho(g)})$ have the same row space and the same column space.
        In fact, the left (resp.\ right) factor $\Gamma(\ov{x^r})$ simply permutes the rows (resp.\ columns) of $\Gamma(\phi(\ov{g}))$.

\item If~$g$ is a divisor of~$x^n-1$, then so is~$\rho(g)$.
        But the representative of $\phi(\ov{g})$ of degree less than~$n$ is in general not a divisor of $x^n-1$.
       Thus, while involution $\ov{g(x)}\mapsto \ov{g(x^{n-1})}$ is the appropriate map for transposition of circulants, it does not
       behave well when it comes to divisors of~$x^n-1$.
        \end{alphalist}
\end{rem}

Now we review the basic algebraic properties of cyclic codes in the terminology of circulant matrices; for further details see for
instance~\cite[Sec.~4.1, 4.2]{HP03}.

\begin{rem}\label{R-Rcomm}
Let $x^n-1=hg$, where $g=\sum_{i=0}^rg_ix^i,\,h=\sum_{i=0}^k h_i x^i$ are monic of degree~$r$ and~$k$, respectively.
Let~$\cC$ be the cyclic code $\cC=\vv\big((\ov{g})\big)\subseteq\F^n$.
\begin{alphalist}
\item The ideal $(\ov{g})$ has dimension $k:=n-r$ as an $\F$-vector space and $\ov{g},\ldots,\ov{x^{k-1}g}$ is a basis.
         Thanks to the isomorphism~$\vv$, this implies that $\cC$ is the row space of the circulant $\Gamma(g)$ and actually of its first~$k$ rows (see also Remark~\ref{R-MgMh}(b)).
         Since $\deg(g)=r$, these first rows have the form
         \[
              G=\begin{pmatrix}\vv(\ov{g})\\ \vv(\ov{xg})\\ \vdots\\ \vv(\ov{x^{k-1}g})\end{pmatrix}
          =\begin{pmatrix}g_0&g_1        &\cdots       &g_r     &        &   &\\
                          &g_0&g_1&\cdots&g_r&   &\\
                    &           &\ddots     &\ddots   &     &\ddots&\\
                    &           &           &g_0&g_1&\cdots&g_r \end{pmatrix}\in\Mat_{k,n}(\F),
          \]
           which is the well-known \textbf{generator matrix} of the cyclic code generated by~$g$.
\item  Remark~\ref{R-MgMh}(a) yields $\Gamma(\ov{g})\Gamma(\ov{h})=0$ and thus $\Gamma(\ov{g})\Gamma(\phi(\ov{h}))\T=0$
         for $\phi$ as in Remark~\ref{R-MgMh}(c).
         As a consequence, the code $\cC=\vv\big((\ov{g})\big)$ satisfies
         \[
            \cC=\{v\in\F^n\mid \Gamma(\phi(\ov{h}))v\T=0\}.
         \]
         Since $\deg(\phi(h))=\deg(h)=k$, the last~$n-k$ rows of $\Gamma(\phi(\ov{h}))$ form a basis of the row space of this matrix.
         Writing $h=\sum_{i=0}^k h_i x^i$, all of this implies $\cC=\{v\in\F^n\mid Hv\T=0\}$, where
         \[
              H
          =\begin{pmatrix}h_k&h_{k-1}        &\cdots       &h_0     &        &   &\\
                          &h_k&h_{k-1}&\cdots&h_0&   &\\
                    &           &\ddots     &\ddots   &     &\ddots&\\
                    &           &           &h_k&h_{k-1}&\cdots&h_0 \end{pmatrix}\in\Mat_{n-k,n}(\F),
          \]
           which is known as the \textbf{parity check matrix} of~$\cC$. This is also the submatrix consisting of the first~$n-k$ rows  of $\Gamma(\ov{\rho(h)})$, where~$\rho(h)$ is the reciprocal of~$h$.
\item $\rho(h)\rho(g)=x^n-1$, and the dual code~$\cC^\perp$ (see \cite[Sec~1.3]{HP03})
          is cyclic with generator and parity check polynomial $\rho(h)/h_0$ and $\rho(g)/g_0$, respectively.
 \end{alphalist}
\end{rem}

\section{Algebraic Theory of Skew-Cyclic Codes with General Modulus}\label{S-BSCC}
In this section we introduce the notion of skew-cyclic codes in most generality and present the basic algebraic properties.
Later we will restrict ourselves to more special cases.
The material of this section is drawn from \cite{BoUl09a,BoLe13,FGL15}. We will put an emphasis on the generalization of the circulant.

From now on we consider the skew-polynomial ring $\Fxs$ where $\F=\F_{q^m}$ and $\sigma$ is the $q$-Frobenius.
In order to generalize the last section we need to first generalize the quotient ring $\cR=\F[x]/(x^n-1)$.
To do so, note that for any $f\in\Fxs$ we obtain a left $\Fxs$-module $\Fxs/\lideal{f}$ (we may, of course, also consider right ideals and right modules).
It is known from basic module theory that this module is a ring if and only if~$f$ is a two-sided polynomial; see Remark~\ref{R-TwoSided}.
In this section, no ring structure is needed and thus we fix the following setting until further notice.

Let $f\in\Fxs$ be a monic polynomial of degree~$n$, which we call the \textbf{modulus}, and consider the left $\Fxs$-module
\[
    \cR_f=\Fxs/\lideal{f}.
\]
Note that the left module structure means that $z\ov{g}=\ov{zg}$ for any $z,g\in\Fxs$, where $\ov{g}$ denotes the coset $g+\lideal{f}$ in~$\cR_f$.
As in the previous section we consider the map
\[
  \pp_f:\,\F^n\longrightarrow \cR_f,\ (c_0,\dots, c_{n-1})\longmapsto \ov{\sum_{i=0}^{n-1} c_i x^i}.
\]
It is crucial that the coefficients~$c_i$ appear on the left of~$x$, because this turns~$\pp_f$ into an
isomorphism of (left) $\F$-vector spaces.
This map will relate codes in~$\F^n$ to submodules in~$\cR_f$.
Again we set $\vv_f=\pp_f^{-1}$.
The map~$\vv_f$ coincides with the map $\phi$ given in \cite[Prop.~3]{BoLe13}, where it is defined with the aid of a semi-linear map based on the companion matrix of~$f$;
see~\eqref{e-CompMat}.

The following facts about submodules of~$\cR_f$ are straightforward generalizations of the commutative case,
see \cite[Thm.~4.2.1]{HP03},
and are proven in exactly the same way (with the aid of right division with remainder in $\Fxs$).
We use the notation $\lideal{\ov{g}}$ for the left submodule $\{z\ov{g}\mid z\in\Fxs\}$ of $\cR_f$
\index{$\lideal{\ov{g}}$: Left submodule generated by $\ov{g}$}
generated by~$\ov{g}$.

\begin{prop}\label{P-submoduleS}
Let~$M$ be a left submodule of~$\cR_f$.
Then~$M=\lideal{\ov{g}}$, where~$g\in\Fxs$ is the unique monic polynomial of smallest degree such that
      $\ov{g}\in M$.
      Alternatively, $g$ is the unique monic right divisor of~$f$ such that $\lideal{\ov{g}}=M$.
      Finally $g\rmid h$ for any $h\in\Fxs$ such that $\ov{h}\in M$.
\end{prop}

The following definition of skew-cyclic codes was first cast in~\cite{BGU07} for the case where~$f$ is a central polynomial of the form $f=x^n-1$, i.e., $\sigma^n=\id$.
In the form below the definition appeared in~\cite{BoUl09a}.
A different, yet equivalent, definition was cast in \cite[Def.~3]{BoLe13}.
Note the generality of the setting.
It includes for instance the case $f=x^n$, for which even in the commutative case the terminology `cyclic' may be questionable because reduction modulo~$f=x^n$
simply means truncating the given polynomial at power~$x^{n}$.
Yet, the basic part of the algebraic theory, presented in this section, applies indeed to this generality, and only in the next section we will restrict ourselves to skew-constacyclic codes.

\begin{defi}\label{D-SCC}\index{$(\sigma,f)$-skew-cyclic}\index{Skew-cyclic}\index{Skew-constacyclic}
\index{$\sigma$-cylic}
A subspace $\cC\subseteq\F^n$ is called $(\sigma,f)$-\textbf{skew-cyclic} if
$\pp_f(\cC)$ is a submodule of~$\cR_f$.
For $a\in\F^*$ the code~$\cC\subseteq\F^n$ is called $(\sigma,a)$-\textbf{skew-constacyclic} if it is $(\sigma,x^n-a)$-skew-cyclic.
The code is called $\sigma$-\textbf{cyclic} if it is $(\sigma,1)$-skew-constacyclic.
We will also call the image~$\pp_f(\cC)$ a cyclic code (of the same type).
\end{defi}

Thus, up to the isomorphism~$\pp_f$ the skew-cyclic codes are the submodules of~$\cR_f$.
In the literature, the above defined codes are often called \emph{ideal~$\sigma$-codes} if~$f$ generates a two-sided ideal and \emph{module-$\sigma$-codes} otherwise.
The $q$-cyclic codes introduced in \cite{Gab09} are the $(\sigma,1)$-skew-constacyclic codes for the case where $m=n$.

Skew-constacyclic codes can easily be described in~$\F^n$.
Just as in the commutative case one observes that  a subspace~$\cC\subseteq\F^n$ is
$(\sigma,a)$-skew-constacyclic if and only if
\begin{equation}\label{e-shift}
   (c_0,\ldots, c_{n-1})\in\cC\Longrightarrow (a\sigma(c_{n-1}),\sigma(c_0),\cdots,\sigma(c_{n-2}))\in\cC.
\end{equation}
Indeed, the right hand side is simply $\vv_{x^n-a}(\ov{x\sum_{i=0}^{n-1}c_i x^i})$.
In other words, a $(\sigma,a)$-skew-constacyclic code is invariant under the $\sigma$-semilinear map induced by the companion matrix $C_{x^n-a}$
(see~\eqref{e-CompMat} and the paragraph thereafter).
This characterization also generalizes to  $(\sigma,f)$-skew-cyclic codes; see \cite[pp.~466]{BoLe13}.

Thanks to Proposition~\ref{P-submoduleS} every $(\sigma,f)$ skew-cyclic code is generated by a single element in~$\cR_f$ (the analogue of principal ideals), i.e., has a generator polynomial.
As a consequence, the number of $(\sigma,f)$-skew-cyclic codes equals the number of monic right divisors of~$f$.
This leads in general to a significantly larger number of skew-cyclic codes than classical cyclic codes.
For instance, in the constacyclic case where $f=x^{15}-\omega$ and $\omega\in\F_4$ satisfies $\omega^2+\omega+1=0$, the polynomial~$f$ has~$8$ monic divisors
(including the trivial ones) in the commutative ring $\F_4[x]$, whereas it has~$32$ monic right divisors in the skew-polynomial ring $\F_4[x;\sigma]$.

Proposition~\ref{P-submoduleS} allows us to present generator matrices just as for the commutative case.
Again we will do so with the aid of circulants.
The definition of a skew circulant matrix is straightforward.

\begin{defi}\label{D-CircMat}
For $\ov{g}\in\cR_f$ define the $(\sigma,f)$-\textbf{circulant}\index{ $(\sigma,f)$-circulant}\index{Skew circulant}
\[
    \Gamma_f^{\sigma}(\ov{g}):=\begin{pmatrix}\vv_f(\ov{g})\\ \vv_f(x\ov{g})\\ \vdots \\
                  \vv_f(x^{n-2}\ov{g})\\ \vv_f(x^{n-1}\ov{g})\end{pmatrix}\in\Mat_{n,n}(\F).
\]
In the case where $f=x^n-a,\,a\in\F^*$, we write $\Gamma_a^{\sigma}$ instead of $\Gamma_{x^n-a}^{\sigma}$.
We call any matrix of the form $\Gamma_f^{\sigma}(\ov{g})$ a \textbf{skew circulant}.
\end{defi}

One may regard $\Gamma_f^{\sigma}(\ov{g})$ as the matrix representation of the left $\F$-linear map on $\cR_f$ given by right multiplication by $\ov{g}$
with respect to the basis $\{\ov{x^0},\ldots,\ov{x^{n-1}}\}$.
If $f=x^n-a$, the skew circulant of~$\ov{g}$ can be given explicitly.
For $g=\sum_{i=0}^{n-1}g_ix^i$ we have $\Gamma_a^{\sigma}(\ov{g})=$

\[
   \begin{pmatrix}g_0\!&\!g_1\!&\!g_2\!&\!\ldots\!&\!g_{n-2}\!&\!g_{n-1}\\
           a\sigma(g_{n-1})\!&\!\sigma(g_0)\!&\!\sigma(g_1)\!&\!\ldots \!&\!\sigma(g_{n-3})\!&\!\sigma(g_{n-2})\\
           a \sigma^2(g_{n-2})\!&\!\sigma(a)\sigma^2(g_{n-1})\!&\!\sigma^2(g_0)\!&\!\ldots \!&\!\sigma^2(g_{n-4})\!&\!\sigma^2(g_{n-3})\\
                          \vdots \!&\!\vdots \!&\!\ddots  \!& &\!\vdots\!&\!\vdots \\
           a\sigma^{n-2}(g_2)\!&\!\sigma(a)\sigma^{n-2}(g_3)\!&\!\sigma^2(a)\sigma^{n-2}(g_4)\!&\!\ldots\!&\!\sigma^{n-2}(g_0)
                                \!&\!\sigma^{n-2}(g_1)\\
          a\sigma^{n-1}(g_1)\!&\!\sigma(a)\sigma^{n-1}(g_2)\!&\!\sigma^2(a)\sigma^{n-1}(g_3)\!&\!\ldots\!&\!
                                 \sigma^{n-2}(a)\sigma^{n-1}(g_{n-1})\!&\!\sigma^{n-1}(g_0)
      \end{pmatrix}.
\]
If $f=x^n-1$, then $ \Gamma_1^{\sigma}(\ov{g})=D_g$, the Dickson matrix in~\eqref{e-Dickson}, and this specializes to the classical circulant~$\Gamma(g)$ in~\eqref{e-Circg}
if $\sigma=\id$.
Furthermore, for any monic~$f$ the skew circulant $\Gamma_f^{\sigma}(\ov{x})$ equals~$C_f$, the companion matrix of~$f$ in~\eqref{e-CompMat}.
Note also that modulo $x^n-a$
\begin{equation}\label{e-Gammax}
  \Gamma_a^{\sigma}(\ov{x})=\begin{pmatrix}   &1& & &\\ & &1& & \\ & & & \ddots & \\ & & & &1\\ a& & & &\end{pmatrix}\ \text{ and }\
  \Gamma_a^{\sigma}(\ov{x^2})=\begin{pmatrix}  & &\!\!1& & \\ & & &\ddots &\\ & & & &1\\ a& & & &\\ &\!\!\sigma(a)\!\!& & &\end{pmatrix}.
\end{equation}

\begin{exa}\label{E-CircSkew}
Let $f=x^7+\alpha\in\F_8[x;\sigma]$, where $\alpha^3+\alpha+1=0$ and $\sigma$ is the 2-Frobenius. Let $g=x^4+\alpha x^3+\alpha^5x^2+\alpha$.
Then $g$ is a right divisor of~$f$ and
\[
   \Gamma:=\Gamma_f^{\sigma}(\ov{g})=\begin{pmatrix}\alpha&0&\alpha^5&\alpha&1&0&0\\0&\alpha^2&0&\alpha^3&\alpha^2&1&0\\0&0&\alpha^4&0&\alpha^6&\alpha^4&1
             \\\alpha&0&0&\alpha&0&\alpha^5&\alpha\\\alpha^3&\alpha^2&0&0&\alpha^2&0&\alpha^3\\1&\alpha^6&\alpha^4&0&0&\alpha^4&0\\0&1&\alpha^5&\alpha&0&0&\alpha\end{pmatrix}.
\]
The first row is simply the vector of left coefficients of~$g$.
The second and third row of~$\Gamma$ are the cyclic shift of the previous row followed by the map~$\sigma$ applied entrywise.
Thus, the first~$3$ rows do not depend on~$f$.
Only in the last~$4$ rows, where $\deg(x^ig)$ is at least~$7$, reduction modulo $\lideal{f}$ kicks in.
\end{exa}

Now one obtains the straightforward analog of Remark~\ref{R-Rcomm}(a): every $(\sigma,f)$-skew-cyclic code has a
generator matrix that reflects the skew-cyclic structure.
Consider a general modulus~$f$ of degree~$n$.
For any matrix~$G$ we use the notation $\rs(G)$ for the row span of~$G$.\index{$\rs(G)$: Row span of a matrix~$G$}

\begin{prop}[see also \mbox{\cite{BoUl09}, \cite[Cor.~2.4]{FGL15}}]\label{P-Mbasis}
Let $\cM=\lideal{\ov{g}}\subseteq\cR_f$, where $g=\sum_{i=0}^r g_i x^i\in\Fxs$ has degree~$r$.
Then:
\begin{alphalist}
\item For any $u\in\F^n$ we have $\pp_f(u\Gamma_f^{\sigma}(\ov{g}))=\pp_f(u)\ov{g}$.
\item $\vv_f(\cM)=\rs(\Gamma_f^{\sigma}(\ov{g}))$.
\item Suppose that~$g$ is a right divisor of~$f$ of degree~$r$.
        Then~$\cM$ is a left~$\F$-vector space of dimension~$k:=n-r$ with basis $\{\ov{g},\,\ov{xg},$ $\ldots,\,\ov{x^{k-1}g}\}$.
        As a consequence, $\rank\big(\Gamma_f^{\sigma}(\ov{g})\big)=k$ and
        \[
          \vv_f(\cM)=\rs(G),
        \]
        where~$G\in\Mat_{k,n}(\F)$ consists of the first~$k$ rows of the skew circulant $\Gamma_f^{\sigma}(\ov{g})$:
        \begin{equation}\label{e-Gskew}
          G=\begin{pmatrix}\vv_f(\ov{g})\\ \vv_f(\ov{xg})\\ \vdots\\ \vv_f(\ov{x^{k-1}g})\end{pmatrix}
          =\begin{pmatrix}g_0&g_1        &\cdots       &g_r     &        &   &\\
                            &\sigma(g_0)&\sigma(g_1)&\cdots&\sigma(g_r)&   &\\
                            &           &\ddots     &\ddots   &     &\ddots&\\
                            &           &           &\!\!\sigma^{k-1}(g_0)&\!\!\sigma^{k-1}(g_1)&\cdots&\sigma^{k-1}(g_r)
          \end{pmatrix}.
        \end{equation}
        If, in addition,~$g$ is monic, we call it the \textbf{generator polynomial} of the $(\sigma,f)$-skew-cyclic code~$\cM$.
        \index{Generator polynomial of a skew-cyclic code}
\item Let $z\in\Fxs$ and $g=\gcrd(z,f)$. Then $\lideal{\ov{z}}=\lideal{\ov{g}}$ and thus
        $\rs(\Gamma_f^{\sigma}(\ov{z}))=\rs(\Gamma_f^{\sigma}(\ov{g}))$.
\end{alphalist}
\end{prop}

In order to provide a feeling for the line of reasoning we provide a short proof.

\noindent\textbf{Proof:}
(a) For any $u_i\in\F$ we have $(u_0, \ldots, u_{n-1})\Gamma_f^{\sigma}(\ov{g})=\sum_{i=0}^{n-1}u_i\vv_f(\ov{x^ig})=\vv_f\big((\sum_{i=0}^{n-1}u_ix^i)\ov{g}\big)$;
hence $(\sum_{i=0}^{n-1}u_ix^i)\ov{g}=\pp_f((u_0, \ldots, u_{n-1})\Gamma_f^{\sigma}(\ov{g}))$, which proves the statement.
\\
(b)
 `$\supseteq$' follows from~(a).
`$\subseteq$' Consider $\ov{zg}\in\lideal{g}$, where $z\in\Fxs$.
Thanks to Theorem~\ref{R-PropR}(c) there exist $u,v\in\Fxs$ such that $ug=vf=\lclm(g,f)$ and $\deg(u)\leq n$.
Right division with remainder of~$z$ by~$u$ provides us with polynomials $t,r\in\Fxs$ such that $z=tu+r$ and $\deg(r)<\deg(u)\leq n$.
Now we have $\ov{zg}=\ov{tvf+rg}=\ov{rg}$, and writing $r=\sum_{i=0}^{n-1}r_ix^i$, we conclude $\vv_f(\ov{zg})=\vv_f(\ov{rg})=(r_0,\ldots, r_{n-1})\Gamma_f^{\sigma}(\ov{g})$.
\\
(c) Let $hg=f$. It suffices to show that every $\ov{zg}\in\lideal{\ov{g}}$ is of the form $\ov{rg}$, where $\deg{r}<k$.
But this follows from the previous part because $hg=f=\lclm(g,f)$.
\\
(d) $\lideal{\ov{z}}\subseteq\lideal{\ov{g}}$ holds since $g\rmid z$.
For the other containment use a Bezout identity $g=uf+vz$ with $u,v\in\Fxs$ (see Theorem~\ref{R-PropR}(b)) and take cosets.
\hfill$\Box$

As we noticed already in Example~\ref{E-CircSkew}, the matrix~$G$ in~\eqref{e-Gskew} above does not depend on the modulus~$f$.
The dependence materializes only through the fact that the code~$\cM$ is $(\sigma,f)$-skew-cyclic.
As a consequence, a given subspace of~$\F^n$ may be $(\sigma,f)$-skew-cyclic for various moduli~$f$.
This has been studied in further detail in~\cite[Sec.~2]{BoUl09}.
Therein, the authors discuss existence and degree of the smallest monic \emph{two-sided} polynomial $f$ such that
$\ov{f}\in\lideal{\ov{g}}$; such~$f$ is called the bound of~$g$ (see also \cite[Ch.~3]{Jac43}).
Its degree is the shortest length in which the given~$g$ generates an ideal-$\sigma$-code.

In this context we wish to remark that a code~$\cC\neq\F^n$ can only be $(\sigma,a)$-skew-constacyclic with respect to at most one polynomial $x^n-a$.
Indeed, if a polynomial~$g$ is a right divisor of $x^n-a$ and $x^n-b$ in~$\Fxs$, then $a=b$ or~$g\in\F^*$.
However, it is possible that a code $\cC\neq\F^n$ is $(\sigma,a)$-skew-constacyclic and $(\sigma',b)$-skew-constacyclic for some
$\sigma\neq\sigma'$ and $a\neq b$.
For instance, over the field $\F_4$ the polynomial $g=x+\omega$ is a right divisor of $x^2-1$ in $\F_4[x;\sigma]$ and a divisor of $x^2-\omega^2$ in $\F_4[x]$.
Hence $\rs(\omega\ \; 1)\subseteq\F_4^2$ is $(\sigma,1)$-skew-constacyclic and $(\id,\omega^2)$-skew-constacyclic.

Let us now study the map induced by the skew circulant.

\begin{rem}\label{R-PropCircMat}
Consider the map $\Gamma_f^{\sigma}:\;\cR_f\longrightarrow \Mat_{n,n}(\F),\ \ov{g}\longmapsto\Gamma_f^{\sigma}(\ov{g})$.
\begin{alphalist}
\item $\Gamma_f^{\sigma}$ is injective and additive.
\item $\Gamma_f^{\sigma}(c\ov{g})=\Gamma_{f'}^{\sigma}(\ov{c})\Gamma_f^{\sigma}(\ov{g})$ for all $c\in\F,\,g\in\cR_f$ and all monic~$f'\in\Fxs$ of degree~$n$.
      This follows directly from the definition along with the fact that
      \[
          \Gamma_{f'}^{\sigma}(\ov{c})=\begin{pmatrix}c& & & \\ & \sigma(c)& & \\ & &\ddots& \\ & & &\sigma^{n-1}(c)\end{pmatrix}\ \text{ for any monic~$f'$ of degree~$n$}.
      \]
      As a consequence, $\Gamma_f^{\sigma}$ is not $\F$-linear (unless $\sigma=\text{id}_{\F}$), but it is
     $\F_q$-linear (recall that $\F_q$ is the fixed field of~$\sigma$).
\item $\Gamma_f^{\sigma}$ is not multiplicative, that is,
      $\Gamma_f^{\sigma}(\ov{gg'})\neq\Gamma_f^{\sigma}(\ov{g})\Gamma_f^{\sigma}(\ov{g'})$ in general.
      This simply reflects the fact that~$\cR_f$ is not a ring.
\end{alphalist}
\end{rem}

By~(c), the identity $hg=f$ does not imply
$\Gamma_f^{\sigma}(\ov{h})\Gamma_f^{\sigma}(\ov{g})=0$.
The situation becomes much nicer when~$f$ is two-sided.
The following is obtained by applying Proposition~\ref{P-Mbasis}(a) twice to $\pp_f(u\Gamma_f^{\sigma}(\ov{g})\Gamma_f^{\sigma}(\ov{g'}))$ for $u\in\F^n$.

\begin{theo}[see also \mbox{\cite[Thm.~3.6]{FGL15}}]\label{T-fTwoSided}
Let $f\in\Fxs$ be two-sided; thus $\cR_f$ is a ring. Then
\[
   \Gamma_f^{\sigma}(\ov{gg'})=\Gamma_f^{\sigma}(\ov{g})\Gamma_f^{\sigma}(\ov{g'})\text{ for all }g,\,g'\in\Fxs.
\]
Hence~$\Gamma_f^{\sigma}$ is an $\F_q$-algebra isomorphism between~$\cR_f$ and the subring $\Gamma_f^{\sigma}(\cR_f)$ of $\Mat_{n,n}(\F)$ consisting of the $(\sigma,f)$-circulants.
\end{theo}

This result does not generalize if~$f$ is not two-sided.
For instance,~\eqref{e-Gammax} shows that
$\Gamma_a^{\sigma}(\ov{x^2})\neq \big(\Gamma_a^{\sigma}(\ov{x})\big)^2$ if $\sigma(a)\neq a$.

The following consequence for $(\sigma,f)$-skew-constacyclic codes is immediate.
In~ \cite[Cor.~1]{BoLe13} the matrix $\Gamma_f^{\sigma}(\ov{h'})$ appearing below is called the \emph{control matrix} of the
code~$\cC$. This is not to be confused with the parity check matrix to which we will turn later.

\begin{cor}\label{C-fTwoSided}
Let $f\in\Fxs$ be two-sided and $f=hg=gh'$ for some $g,h,h'\in\Fxs$.
Then $\Gamma_f^\sigma(\ov{g})\Gamma_f^{\sigma}(\ov{h'})=0$ and
the code $\cC=\vv_f(\lideal{\ov{g}})=\rs(\Gamma_f^\sigma(\ov{g}))$ is the left kernel of the skew circulant $\Gamma_f^{\sigma}(\ov{h'})$.
\end{cor}

It is not hard to see that actually the two-sidedness of~$f$ along with $f=hg$ implies the existence of~$h'$ such that $f=gh'$.

Now that we have a natural notion of  generator matrix for a skew-cyclic code it remains to discuss whether such a code also has a parity check matrix that reflects the skew-cyclic structure.
We have shown in Remark~\ref{R-Rcomm}(b) that in the commutative case the parity check matrix hinges on two facts: (i) the product of circulants is again a circulant, (ii) the transpose of a circulant is a circulant.
Theorem~\ref{T-fTwoSided} shows that property~(i) carries through to the non-commutative case if the modulus is two-sided
(and thus also to the commutative case for arbitrary moduli~$f$ instead of $x^n-1$).
But if~$f$ is not two-sided, then even in the skew-constacyclic case (i.e., moduli of the form $f=x^n-a$), the product of two $(\sigma,f)$-circulants is not a
$(\sigma,f)$-circulant in general.
However, we will encounter a proxy of such multiplicativity in the next section that fully serves our purposes.

Transposition of $(\sigma,f)$-circulants is an even bigger obstacle.
For general modulus~$f$ the transpose of a $(\sigma,f)$-circulant need not be a $(\sigma',f')$-circulant for any
automorphism~$\sigma'$ and any modulus~$f'$ of the same degree as~$f$.
This is actually not very surprising because even in the commutative case the transpose of a circulant in the sense of Definition~\ref{D-CircMat}
need not be a circulant.
A trivial example is $f=x^n$ and $g=x^r$, but examples also exist for polynomials~$f$ with nonzero constant term.
The following (noncommutative) example illustrates this.

\begin{exa}\label{E-TransNotCirc}
Consider $f=x^3+x^2+\omega^2,\,g=x^2+\omega x+\omega\in\F_4[x;\sigma]$, where $\omega^2+\omega+1=0$ and~$\sigma$ is the $2$-Frobenius.
Then $g$ is a right divisor of~$f$ and
\[
  G:=\Gamma_f^{\sigma}(\ov{g})=\begin{pmatrix}\omega & \omega & 1 \\
                                          \omega^2 & \omega^{2} & \omega \\
                                          \omega &  \omega & 1
                                          \end{pmatrix}.
\]
The matrix~$G$ has rank~$1$ and thus generates a 1-dimensional $(\sigma,f)$-skew-cyclic code $\cC=\vv_f(\lideal{\ov{g}})$.
Suppose the transpose~$G\T$ is a $(\sigma',f')$-circulant for some automorphism~$\sigma'$ and $f'\in\F[x;\sigma']$ of degree~$3$, say $G\T=\Gamma_{f'}^{\sigma'}(\ov{g'})$.
Then clearly~$g'$ is given by the first column of~$G$; hence $g'=\omega+\omega^2x+\omega x^2$.
Using for instance SageMath one checks that $G\T\neq\Gamma_{f'}^{\sigma'}(\ov{g'})$ for any automorphism~$\sigma'$ of~$\F_4$ and any $f'\in\F_4[x;\sigma']$ of degree~$3$ (even non-monic).
Furthermore, there exists no skew circulant $H=\Gamma_{f'}^{\sigma'}(\ov{h})$ such that $\rank(H)=2$ and $GH\T=0$.
This means there is no analogue of Remark~\ref{R-Rcomm}(b),(c):
$\cC$ does not have a skew circulant parity check matrix and~$\cC^\perp$ is not $(\sigma',f')$-skew-cyclic for any $(\sigma',f')$.
\end{exa}

In the next section we restrict ourselves to skew-constacyclic codes and will see that in that case these obstacles can be overcome.

We close this section by presenting a different type of parity check matrix, namely a generalization of the Vandermonde type parity check matrix for classical cyclic codes.
Recall W-polynomials from Section~\ref{S-AlgSets}.
The following result is immediate with the definition of the skew Vandermonde matrix in~\eqref{e-Vand}.

\begin{theo}[\mbox{\cite[Prop.~4]{BoLe13}}]\label{T-VanPCM}
Let $f\in\Fxs$ be  any monic modulus of degree~$n$ and $g\in\Fxs$ be a monic right divisor of~$f$ of degree~$r$.
Suppose~$g$ is a W-polynomial.
Thus  we may write $g=\lclm (x-a_1,\dots,x-a_r)$ for distinct $a_1,\ldots,a_r\in\F$; see Theorem~\ref{T-WPoly}(1)(ii).
Let $M=V_n(a_1,\ldots,a_r)\in\Mat_{n,r}(\F)$ be the skew Vandermonde matrix.
Then the cyclic code $\cC=\vv(\lideal{\ov{g}})$ is given by
\[
   \cC=\{(c_0,\ldots, c_{n-1})\mid (c_0,\ldots, c_{n-1})M=0\}.
\]
\end{theo}

\section{Skew-Constacyclic Codes and their Duals}\label{S-SkewConsta}
We now restrict ourselves to skew-constacyclic codes, that is to modulus $x^n-a$.
In this case we are able to obtain a parity check matrix, and thus a generator matrix of the dual, that reflects the skew-constacyclic structure.
The material is mainly drawn from \cite{BoUl09a,BoUl14,FGL15}.

Throughout, we fix a modulus $f=x^n-a$ for some $a\in\F^*$.
In order to formulate  the main results we need, as in the commutative case, the reciprocal of a polynomial.
In the noncommutative case this can be done in different ways depending on the position of the coefficients.
The following left version of~\eqref{e-Recip} will suffice for this survey. Let
\[
  \rho_l:\Fxs\longrightarrow\Fxs,\quad \sum_{i=0}^r g_i x^i\longmapsto \sum_{i=0}^r x^{r-i}g_i=\sum_{i=0}^r\sigma^i(g_{r-i})x^i \ \text{(where $g_r\neq0$).}
\]
Then $\rho_l(g)$ is called the \textbf{left reciprocal} of~$g$.\index{Skew polynomial!Left reciprocal}
Furthermore, we extend the automorphism~$\sigma$ to the ring $\Fxs$ via $\sigma(\sum_{i=0}^r g_i x^i)=\sum_{i=0}^r \sigma(g_i)x^i$.
Then~$\sigma$ is a ring automorphism of~$\Fxs$ satisfying $xg=\sigma(g)x$ for all $g\in\Fxs$.

The following partial product formula for skew circulants will be sufficient to discuss the duals of skew-constacyclic codes.
Recall that for $f=x^n-a$ we denote the skew circulant $\Gamma_f^\sigma$ by $\Gamma_a^\sigma$.
We will have to deal with different moduli, $x^n-a$ and $x^n-c$, and of course the notation $\Gamma_c^{\sigma}(\ov{g})$ means that the coset of~$g$ is taken modulo $\lideal{x^n-c}$.

\begin{theo}[\mbox{\cite[Thm.~5.3]{FGL15}}]\label{T-ProdSkewCirc}
Let $x^n-a=hg$. Set $c=\sigma^n(g_0)ag_0^{-1}$, where $g_0$ is the constant coefficient of~$g$.
Then $x^n-c=\sigma^n(g)h$ and
\[
   \Gamma_a^{\sigma}(\ov{g'g})=\Gamma_c^\sigma(\ov{g'})\Gamma_a^{\sigma}(\ov{g}) \ \text{ for any }\ g'\in\Fxs.
\]
\end{theo}

Note that~$c$ defined in the theorem is the conjugate~$a^{g_0}$ with respect to the automorphism~$\sigma^n$ in the sense of Definition~\ref{D-Conj}.
If $c=a$ (i.e., $\sigma^n(g_0)=g_0$) we have the much nicer formula
$\Gamma_a^{\sigma}(\ov{g'g})=\Gamma_a^\sigma(\ov{g'})\Gamma_a^{\sigma}(\ov{g})$, which may be regarded as a generalization of the two-sided case in Theorem~\ref{T-fTwoSided}.
However, the above result holds true only for right divisors~$g$ of~$x^n-a$.
Check, for instance, with the aid of~\eqref{e-Gammax} that $\Gamma_a^{\sigma}(\ov{x(x+1)})\neq \Gamma_b^{\sigma}(\ov{x})\Gamma_a^\sigma(\ov{x+1})$ for any $b\neq0$ unless $a=\sigma(a)=b$.

The above product formula plays a central role in the following quite technical result.
It tells us that the transpose of a $(\sigma, x^n-a)$-circulant is a $(\sigma,x^n-a')$-circulant for a suitable constant~$a'$.

\begin{theo}[\mbox{\cite[Thm.~5.6]{FGL15}}]\label{T-CircTrans}
Suppose $x^n-a=hg$ for some $g,h\in\Fxs$ of degree~$r$ and $k=n-r$, respectively.
Set again $c=\sigma^n(g_0)ag_0^{-1}$, where $g_0$ is the constant coefficient of~$g$.
Then
\[
  \Gamma_a^\sigma(\ov{g})\T=\Gamma_{c^{-1}}^\sigma(\ov{g^\#})
  =\Gamma_{\sigma^k(c^{-1})}^\sigma(\ov{g^\circ})\Gamma_{c^{-1}}^{\sigma}(\ov{x^k}),
\]
where $g^\#=a\sigma^k(\rho_l(g))x^k$ and $g^\circ=a \sigma^k(\rho_l(g))$.
Furthermore, $g^\circ$ is a right divisor of the modulus $x^n-\sigma^k(c^{-1})$.
\end{theo}

The result generalizes Remark~\ref{R-MgMh}(c) and~(d): if $f=x^n-1=hg$ and $\sigma=\id$, then $c=1$, $g^\circ=\rho(g)$ and thus $g^\#=\rho(g)x^k$.
Furthermore, in general and analogously to Remark~\ref{R-MgMh}(e),~$g^\circ$ is a right divisor of the modulus $x^n-\sigma^k(c^{-1})$, whereas
the representative of $\ov{g^\#}$ of degree less than~$n$ is in general not a divisor of~$x^n-c^{-1}$.
This is the reason why we provide two formulas pertaining to the transpose of a skew circulant $\Gamma_a^{\sigma}(\ov{g})$.
The first one above is interesting in itself as it tells us that the transpose is again a skew circulant.
The second formula says that the skew-constacyclic code $\lideal{\ov{g}}$, i.e., the row space
of~$\Gamma_a^{\sigma}(\ov{g})$, equals the row space of a transposed skew circulant where the representing polynomial is a right divisor of the modulus.
In all these cases it is crucial that~$g$ is a right divisor of the modulus~$x^n-a$ for otherwise the transpose of $ \Gamma_a^\sigma(\ov{g})$ is not a skew circulant in general~\cite[Ex.~5.7]{FGL15}.

Now we are ready to derive a parity check matrix reflecting the skew-constacyclic structure of the code.
The second part of the following theorem appeared first, proven differently, in
\cite[Thm.~8]{BoUl09a}.
The mere $(\sigma,a^{-1})$-skew-constacyclicity of~$\cC^{\perp}$ can also be shown directly with the aid of~\eqref{e-shift}; see \cite[Thm.~2.4]{VaTi18a}.

\begin{theo}[\mbox{\cite[Cor.~4.4, Thm.~5.8 and Thm.~6.1]{FGL15}}]\label{T-PCMDual}
Let $x^n-a=hg$, where $\deg(g)=r$ and $\deg(h)=k=n-r$. Set  $h^{\circ}:=\rho_l(\sigma^{-n}(h))$.
Then $h^\circ\rmid (x^n-a^{-1})$.
Consider the $(\sigma,a)$-skew-constacyclic code $\cC=\vv_{x^n-a}(\lideal{\ov{g}})$.
Then $\Gamma_a^{\sigma}(\ov{g})\Gamma_{a^{-1}}^\sigma(\ov{h^\circ})\T=0$ and
$\rank(\Gamma_{a^{-1}}^\sigma(\ov{h^\circ}))=n-k$.
Hence
\[
    \cC=\rs(\Gamma_a^{\sigma}(\ov{g}))=\{c\in\F^n\mid \Gamma_{a^{-1}}^\sigma(\ov{h^\circ})c\T=0\},
\]
and the $(n-k)\times n$-submatrix consisting of the first $n-k$ rows of $ \Gamma_{a^{-1}}^\sigma(\ov{h^\circ})$ is a parity check matrix of~$\cC$.
As a consequence, the dual code~$\cC^{\perp}$ is $(\sigma,a^{-1})$-skew-constacyclic with (non-monic) generator polynomial $h^\circ$ and generator matrix and parity check matrix
given by the first $n-k$ rows of $ \Gamma_{a^{-1}}^\sigma(\ov{h^\circ})$ and the first~$k$ rows of $\Gamma_a^{\sigma}(\ov{g})$, respectively.
\end{theo}

Clearly, this parity check matrix of~$\cC$ has a form analogous to~\eqref{e-Gskew} and thus reflects the skew-constacyclic structure of~$\cC$.
As in Proposition~\ref{P-Mbasis} its row space equals the row space of the entire skew circulant $\Gamma_{a^{-1}}^\sigma(\ov{h^\circ})$.
In the commutative cyclic case, where $x^n-1=hg=gh$,  we have $a^{-1}=a=1$ and $h^\circ=\rho(h)$ and thus recover Remark~\ref{R-Rcomm}(b).

Having an understanding of the dual of skew-constacyclic codes, we can address self-duality.
The following corollary is immediate.

\begin{cor}[\mbox{\cite[Prop.~13]{BoUl09a}, \cite[Cor.~6]{BoUl14a}, \cite[Cor.~6.2]{FGL15}}]\label{C-SelfDual}
If there exists a $(\sigma,a)$-skew-consta\-cyclic self-dual code of length~$n$, then~$n$ is even and $a=\pm1$.
More specifically, let~$n$ be even and  consider the modulus $x^n-\epsilon$, where $\epsilon\in\{1,-1\}$.
Then there exists a self-dual skew-constacyclic code of length~$n$ if and only if there exists a polynomial~$h\in\Fxs$ such that
$x^n-\epsilon=h h^\circ$.
In this case the self-dual code is given by $\cC=\vv_{x^n-\epsilon}(\lideal{\ov{h^\circ}})$.
\end{cor}

In~\cite{BoUl14a,Bou16} the identity $x^n-\epsilon=h h^\circ$ is exploited to enumerate or construct self-dual skew-constacyclic codes, some with very good minimum distance, e.g.\ \cite[Ex.~30]{BoUl14a}.

Let us briefly turn to the notion of check polynomials for skew-constacyclic codes.
Recall that in the classical case, where $x^n-1=hg$ in $\F[x]$ we call~$h$ a check polynomial for the simple reason that
$\ov{z}\in\ideal{\ov{g}}\Leftrightarrow \ov{zh}=0$ for any $z\in\F[x]$.
In other words $\ideal{\ov{h}}$ is the annihilator ideal of $\ideal{\ov{g}}$.
The following generalization to $\Fxs$ is based on the fact~\cite[Thm.~4.2]{FGL15} that $x^n-a=hg$ implies $x^n-\tilde{c}=g\sigma^{-n}(h)$ for~$\tilde{c}$ defined below.

\begin{theo}[\mbox{\cite[Prop.~6.5]{FGL15}}]\label{T-CheckPoly}
Let $x^n-a=hg$ and set $c=\sigma^n(g_0)ag_0^{-1}$, where~$g_0$ is the constant coefficient of~$g$.
Define $\tilde{c}=\sigma^{-n}(c)$.
Then the map
\[
   \Psi:\Fxs/\lideal{x^n-a}\longrightarrow\Fxs/\lideal{x^n-\tilde{c}},\quad
        \ov{z}\longmapsto \ov{z\sigma^{-n}(h)}
\]
is a well-defined left $\Fxs$-linear map with kernel $\lideal{\ov{g}}$.
Therefore we call $\sigma^{-n}(h)$ the \textbf{check polynomial} of the code~$\cC=\vv_{x^n-a}(\lideal{\ov{g}})$.
\index{Check polynomial of a skew-cyclic code}
\end{theo}

One has to be aware that the check equation $\ov{z\sigma^{-n}(h)}=0$ has to be carried out modulo $x^n-\tilde{c}$.
The above generalizes \cite[Thm.~2.1(iii)]{GSF16}, where the modulus is central of the form $x^n-1$.
In that case we have $\sigma^n=\id$ and $c=a$, and the above also reflects Corollary~\ref{C-fTwoSided}.
Theorem~\ref{T-CheckPoly} also extends \cite[Lem.~8]{BoUl09}, where general two-sided moduli are considered.

We close this section by mentioning idempotent generators of skew-constacyclic codes.
In \cite{GSF16} the authors consider central moduli of the form $x^n-1$.
Such polynomials have a factorization into pairwise coprime \emph{two-sided maximal} polynomials \cite[Thm.~1.2.17']{Jac96}, which in turn gives rise to a decomposition of
$\Fxs/\ideal{x^n-1}$ into a direct product of rings generated by central idempotents~\cite[Thm.~2.11]{GSF16}.
As a  consequence, just as in the classical cyclic case, a code $\ideal{\ov{g}}$, where~$g$ is a central divisor of $x^n-1$, has a unique central generating idempotent
\cite[Thm.~6.2.15]{Fog16}.
In the thesis~\cite{Fog16} a partial generalization to the non-central case is presented along with the obstacles that occur in this scenario; see \cite[Ch.~6]{Fog16}.

\section{The Minimum Distance of Skew-Cyclic Codes}\label{S-MinDist}
In this section we report on constructions of skew-cyclic codes with designed minimum distance.
We only consider the Hamming distance (there also exist a few results on the rank distance in the literature).
The results are from the papers \cite{BoUl14,GLNN18,TCTi17}.

Throughout, $\F=\F_{q^m}$ and $\sigma$ is the $q$-Frobenius.
Furthermore, we consider a code
\[
  \cC=\vv_f(\lideal{\ov{g}})\ \text{ for some monic $f,g\in\Fxs$ such that $\deg(f)=n$ and $g\rmid f$.}
\]
In the results below we present conditions on~$f,g$ that guarantee a desired minimum distance.
In all interesting cases the generator~$g$ of the code in question will be the least common left multiple of linear factors over some extension field, and thus~$g$ is a W-polynomial over that extension field; see Theorem~\ref{T-WPoly}(1).
In Theorem~\ref{T-VanPCM} we presented a parity check matrix of a skew-cyclic code generated by a W-polynomial
in form of a skew Vandermonde matrix.
This matrix is the basis of the distance results in this section.

The first two results lead to what we will call skew-BCH codes of the first kind.
They are based on generator polynomials with roots that are consecutive ordinary powers of some element.
Thereafter, we present skew-BCH codes of the second kind, which are based on generator polynomials with roots
that are consecutive $q$-powers of some element.
We conclude with two examples illustrating the constructions.

We start with skew-BCH codes of the first kind.

\begin{theo}[\mbox{\cite[Thm.~4]{BoUl14}}]\label{T-BoUl14Thm5}
Fix $b,\delta\in\N$. Suppose there exists some $\alpha\in\ov{\F}$ (the algebraic closure of~$\F$) such that
\[
   \alpha^{\bracket{0}},\alpha^{\bracket{1}},\ldots,\alpha^{\bracket{n-1}}\text{ are distinct and
   $g(\alpha^{b+i})=0$ for $i=0,\ldots,\delta-2$.}
\]
Then the code $\cC=\vv_f(\lideal{\ov{g}})$ has  minimum distance at least~$\delta$.
If~$g$ is the smallest degree monic polynomial with roots $\alpha^b,\ldots,\alpha^{b+\delta-2}$,
then~$\cC$ is called an
$(n,q^m,\alpha,b,\delta)$-\textbf{skew-BCH code of the first kind}. \index{Skew-BCH code of the first kind}
\end{theo}

\begin{cor}[\mbox{\cite[Thm.~5]{BoUl14}}]\label{C-BoUl14Thm5}
Consider the situation of Theorem~\ref{T-BoUl14Thm5} and where $\alpha\in\F$.
Then the polynomial $g':=\lclm(x-\alpha^b,\ldots,x-\alpha^{b+\delta-2})$ is in $\Fxs$ and of degree~$\delta-1$.
Thus for any left multiple~$f'$ of~$g'$ of degree~$n$, the skew-cyclic code $\vv_{f'}(\lideal{\ov{g'}})$ has dimension $n-\delta+1$ and is MDS.
It is called an $(n,q^m,\alpha,b,\delta)$-\textbf{skew-RS code of the first kind}.\index{Skew-RS code of the first kind}
\end{cor}

Theorem~\ref{T-BoUl14Thm5} follows immediately from the fact that the code is contained in the left kernel of the skew Vandermonde matrix  (see~\eqref{e-Vand})
\[
  V_n(\alpha^b,\ldots,\alpha^{b+\delta-2})
    =\begin{pmatrix}1&\cdots&1\\ (\alpha^{\bracket{1}})^b&\cdots&(\alpha^{\bracket{1}})^{b+\delta-2}\\ \vdots& &\vdots\\
         (\alpha^{\bracket{n-1}})^b&\cdots&(\alpha^{\bracket{n-1}})^{b+\delta-2}\end{pmatrix}.
\]
This columns of this matrix consist of consecutive $\bracket{i}$-powers of $\alpha^b,\ldots,\alpha^{b+\delta-2}$ (which simply accounts for skew-polynomial evaluation), while the rows consist of ordinary  consecutive powers of $\alpha^{\bracket{0}},\ldots,\alpha^{\bracket{n-1}}$.
The latter together with the fact that $\alpha^{\bracket{0}},\alpha^{\bracket{1}},\ldots,\alpha^{\bracket{n-1}}$ are distinct, guarantees that any $(\delta-1)\times(\delta-1)$-minor of $V_n(\alpha^b,\ldots,\alpha^{b+\delta-2})$ is nonzero, which establishes the stated designed distance.

There exist some precursors of Theorem~\ref{T-BoUl14Thm5}.
In \cite[Prop.~2]{BGU07} the case where  $f=x^m-1$ (thus $n=m$), $q=2,\,b=0$, and $\alpha$ is a primitive element of~$\F$
was considered.
It was the first appearance of skew-BCH codes.
Subsequently, in \cite[Prop.~2]{CLU09} the modulus~$f$ was relaxed to a two-sided polynomial of degree~$n$, and~$\alpha$ to a primitive element of a field
extension $\F_{q^{s}}$, where $n\leq (q-1)s$.
In the same paper, examples of such codes were constructed by translating the above situation into the realm of linearized polynomials.

Theorem~\ref{T-BoUl14Thm5} has been generalized to the following form.

\begin{theo}[\mbox{\cite[Thm.~4.10]{TCTi17}}]\label{T-TCTi17Thm410}
Let~$f$ have a nonzero constant coefficient.
Suppose there exist $\delta,\,t_1,\,t_2\in\N$ and $b,\,\nu\in\N_0$ and some $\alpha\in\ov{\F}$ such that
\begin{romanlist}
\item $g(\alpha^{b+t_1i+t_2j})=0$ for $i=0,\ldots,\delta-2,\,j=0,\ldots,\nu$,
\item $(\alpha^{t_\ell})^{\bracket{i}}\neq1$ for $i=1,\ldots,n-1,\,\ell=1,2$ (if $\nu=0$, the condition on $\alpha^{t_2}$ is omitted).
\end{romanlist}
Then the code $\vv_f(\lideal{\ov{g}})\subseteq\F^n$ has minimum distance at least~$\delta+\nu$.
It may also be called an $(n,q^m,\alpha,b,t_1,t_2,\delta)$-\textbf{skew-BCH code of the first kind}.
\end{theo}

Note that for $\nu=0$ and $t_1=1$ this result reduces to Theorem~\ref{T-BoUl14Thm5} because Condition~(ii) is equivalent to
$ \alpha^{\bracket{0}},\alpha^{\bracket{1}},\ldots,\alpha^{\bracket{n-1}}$ being distinct.

Since the constructed code has dimension $n-\deg(g)$, it remains to investigate how to find the smallest degree monic polynomial~$g$ satisfying~(i) above (and has degree at most~$n$).
Recall from Proposition~\ref{P-Mbasis}(c) that the modulus~$f$ does not play a role in the generator matrix of the resulting code $\vv_f(\lideal{\ov{g}})\subseteq\F^n$.
Thus, once~$g$ is found, any monic left multiple~$f$ of degree~$n$ suffices.
The polynomial~$g$ is obtained as follows, which is a direct consequence of Example~\ref{E-MinPolya}.

\begin{rem}\label{R-Findg1}
Consider the situation of Theorem~\ref{T-TCTi17Thm410} and suppose $\alpha$ is in the field extension
$\F_{q^{ms}}$ of~$\F=\F_{q^m}$.
Set $T=\{b+t_1i+t_2j\mid \,i=0,\ldots,\delta-2,\,j=0,\ldots,\nu\}$ and
$A=\{\tau(\alpha^{t})\mid \tau\in\Aut(\F_{q^{ms}}\mid\F),\,t\in T\}$.
Then $g:=m_A$ is in $\Fxs$ and is the smallest degree monic polynomial satisfying~(i)  of Theorem~\ref{T-TCTi17Thm410}.
\end{rem}

\begin{exa}\label{E-BCH1}
Consider the field extension $\F_{2^{12}}\mid\F_{2^6}$.
Let $\alpha$ be the primitive element of~$\F_{2^{12}}$ with minimal polynomial $x^{12} + x^7 + x^6 + x^5 + x^3 + x + 1$ and let
$\gamma=\alpha^{65}$, which is thus a primitive element of~$\F_{2^6}$.
As always, $\sigma$ is the $2$-Frobenius.
Let $b=0, t_1=23,\,t_2=1,\,\delta=4,\,\nu=0$.
Since $\nu=0$, Condition~(ii) of Theorem~\ref{T-TCTi17Thm410} amounts to $(\alpha^3)^{\bracket{i}}\neq1$ for $i=0,\ldots,n-1$.
Since $i=12$ is the smallest positive integer such that $(\alpha^3)^{\bracket{i}}=1$, we can construct skew-BCH codes up to length~$12$.
Condition~(i) and Remark~\ref{R-Findg1} shows that the desired~$g$ is given by $g=m_A$, where
\[
    A=\{\alpha^{0},\alpha^{23},\alpha^{46},(\alpha^{0})^{2^6},(\alpha^{23})^{2^6},(\alpha^{46})^{2^6}\}\subseteq\F_{2^{12}}.
\]
Hence~$g=\lclm(x-a\mid a\in A)$, and this results in $ g=x^3+\gamma^{47}x^2+\gamma^{19}x+ \gamma^{40}$.
By construction, for any monic left multiple~$f$ of~$g$ of degree $3\leq n\leq12$, the skew-BCH code
$\cC=\vv_f(\lideal{g})\subseteq\F^n$ has minimum distance at least~$4$ and dimension $n-3$; thus it is MDS.
It is interesting to observe that the polynomial $f=x^{12}-1$ is a left multiple of~$g$, and therefore for length $n=12$
the code is $\sigma$-cyclic.
The code is not $\sigma$-constacyclic for any length between $3\leq n\leq11$.
Finally, note that by definition, $g$ is a W-polynomial in $\F_{2^{12}}[x;\sigma]$; it is, however, not a W-polynomial in
$\F_{2^6}[x;\sigma]$  (it is not the minimal polynomial of its vanishing set in~$\F_{2^6}$).
\end{exa}

We now turn to skew-BCH codes of the second kind.
The codes presented next are $\sigma$-cyclic, i.e., skew-cyclic with respect to the central modulus $x^n-1$, and a priori defined over a field
extension~$\F_{q^n}$ of~$\F=\F_{q^m}$.
The result generalizes the Hartmann-Tzeng Bound for classical cyclic codes.

\begin{theo}[\mbox{\cite[Thm.~3.3]{GLNN18} and \cite[Cor.~5]{MP17} for an earlier, slightly different, version}]\label{T-GLNN18Thm33}
Let $\sigma$ be the $q$-Frobenius on $\F_{q^n}$.
Let $f=x^n-1$ and~$g\in\F_{q^n}[x;\sigma]$ be a right divisor of~$f$.
Let $\alpha\in\F_{q^n}$ be such that $\alpha,\alpha^q,\ldots,\alpha^{q^{n-1}}$ is a normal basis of~$\F_{q^n}$ over~$\F_q$
and set $\beta=\alpha^{-1}\sigma(\alpha)=\alpha^{q-1}$.
Suppose there exist $\delta,\,t_1,\,t_2\in\N$ and $b,\nu\in\N_0$ such that $\gcd(n,t_1)=1$ and $\gcd(n,t_2)<\delta$ and
\[
   g(\beta^{q^{b+it_1+jt_2}})=0\text{ for }i=0,\ldots,\delta-2,\ j=0,\ldots,\nu.
\]
Then the code $\vv_f(\lideal{\ov{g}})\subseteq\F_{q^n}^n$ has minimum distance at least $\delta+\nu$.
\end{theo}

The difference between the versions in~\cite{GLNN18} and~\cite{MP17} is spelled out in \cite[Rem.~A.6]{GLNN18}.

We have seen already in Example~\ref{E-NormBasisVand} that
$x^n-1=\lclm(x-\beta^{q^t}\mid t=0,\ldots,n-1)$.
Therefore the root condition on~$g$ does not clash with the condition that~$g$ be a right divisor of~$f$.

Let us now assume that $n=ms$ so that $\F_{q^n}$ is a  field extension  of $\F_{q^{m}}$.
In~\cite{GLNN18} it is shown how to obtain from the code of the previous theorem a code over
the subfield~$\F=\F_{q^m}$ with the same designed minimum distance  $\delta+s$.
Thus, as for skew-BCH codes of the first kind we want to find the smallest degree monic polynomial~$g$ in $\F_{q^m}[x;\sigma]$ with the desired roots.
This can again be achieved with the aid of Remark~\ref{R-Findg1}, where we simply have to replace the set~$T$ by
$\tilde{T}=\{q^{b+t_1i+t_2j}\mid \,i=0,\ldots,\delta-2,\,j=0,\ldots,\nu\}$.
Noting that $\Aut(\F_{q^{ms}}\mid \F_{q^m})=\{\tau_0,\ldots,\tau_{s-1}\}$, where $\tau_\ell(a)=a^{q^{\ell m}}$, we conclude that the set
$A=\{\tau(\alpha^t)\mid t\in\tilde{T},\tau\in\Aut(\F_{q^{ms}}\mid \F_{q^m})\}$ is given by
\[
   A=\{\alpha^{q^{b+t_1i+t_2j+\ell m}}\mid i=0,\ldots,\delta-2,\,j=0,\ldots,\nu,\,\ell=0,\ldots,s-1\}.
\]
All of this can simply be described in terms of the $q$-exponents within the cyclic group $C_{ms}$ of order $ms$.
Consider~$C_s$, the cyclic group of order~$s$, as a subgroup of~$C_{ms}$.
Furthermore, let $X_0=C_s,X_1,\ldots,X_{m-1}$ be the cosets of~$C_s$ in~$C_{ms}$.
Then Remark~\ref{R-Findg1} and Theorem~\ref{T-GLNN18Thm33} lead to the following.

\begin{theo}[\mbox{\cite[Thm.~4.5]{GLNN18}}]\label{T-GLNN18Thm45}
Let $n=ms$ and consider the situation of Theorem~\ref{T-GLNN18Thm33}.
Consider the set $S=\{b+it_1+jt_2\mid i=0,\ldots,\delta-2,\ j=0,\ldots,\nu\}$ as a subset of~$C_{ms}$ (this is well-defined since $\sigma^{ms}=\id$).
Define $\ov{S}$ as the smallest union of cosets~$X_i$ containing~$S$.
Then the polynomial $g'=\lclm(x-\beta^{q^t}\mid t\in\ov{S})$ is in $\F[x;\sigma]$.
Thus it defines a $(\sigma,f)$-skew cyclic code $\cC=\vv_f(\lideal{\ov{g'}})$ of length~$n=ms$ over~$\F$.
The code~$\cC$ has minimum distance at least $\delta+\nu$ and is called an
$(n,q^m,\alpha,b,t_1,t_2,\delta)$-\textbf{skew-BCH code of the second kind}.\index{Skew-BCH code of the second kind}
\end{theo}

The last part follows from the fact that $g'$ is a left multiple of~$g$ from Theorem~\ref{T-GLNN18Thm33} and thus generates a code
contained in the code from that theorem.

There is a connection between the above and $q$-cyclotomic spaces defined in \cite[Sec.~3.2]{MP17}.
In particular, the $q$-polynomial of an element~$\beta$ defined in \cite[Lem.~3]{MP17} is 
the linearized version of the $\sigma$-minimal polynomial of~$\beta$ as discussed earlier in Example~\ref{E-MinPolya}.
Further details on the connection are given in \cite[Prop.~A.7]{GLNN18}.

\begin{exa}\label{E-SkewBCH}
As in Example~\ref{E-BCH1} consider $\F_{2^{12}}\mid\F_{2^6}$ with the same primitive element $\alpha$ and the same data
$\gamma=\alpha^{65},b=0,t_1=23,t_2=1,\delta=4,\nu=0$.
The element $\alpha^5$ generates a normal basis of $\F_{2^{12}}$ over~$\F_2$.
Thus, $\beta=\alpha^{-5}\sigma(\alpha^5)=\alpha^5$.
We have to consider the set $S=\{b+it_1\mid i=0,1,2\}=\{0,11,10\}$ and
find the smallest union of cosets of $C_2$ in~$C_{12}$ containing~$S$.
This is given by $\ov{S}=\{0,6,11,5,10,4\}$.
Then $\lclm(x-(\alpha^5)^{q^t}\mid t\in\ov{S})=
x^6+\gamma^{61}x^5+\gamma^{41}x^4+\gamma^4x^3+\gamma^{20}x^2+\gamma^{46}x+\gamma^{7}\in\F_{2^6}[x;\sigma]$
generates a skew-cyclic code over $\F_{2^6}$ of length~$12$ and designed minimum distance~$4$.
It has dimension~$6$ and its actual minimum distance is~$6$. Thus the code is not MDS.
\end{exa}

To our knowledge no general comparison of the two kinds of skew-BCH codes has been conducted so far.

Finally, we briefly address evaluation codes in the skew polynomial setting.
Recall that the evaluation below, $p(\alpha_i)$, is carried out according to Definition~\ref{D-Evalf}.

\begin{theo}[\mbox{\cite[Prop.~2]{BoUl14}}]\label{T-BoUl14Prop2}
Let $k\in\{1,\ldots,n-1\}$ and $\alpha_1,\ldots,\alpha_n\in\F$ be such that the skew-Van\-dermonde matrix $V_n(\alpha_1,\ldots,\alpha_n)\in\Mat_{n,n}(\F)$ has rank~$n$.
Then the code
\[
  \cE_{\sigma,\alpha_1,\ldots,\alpha_n}:=\{(p(\alpha_1),\ldots,p(\alpha_n))\mid p\in\Fxs,\,\deg p\leq k-1\}\subseteq\F^n
\]
has dimension~$k$ and minimum distance~$n-k+1$, thus is MDS.
\end{theo}

By Theorem~\ref{T-VandRank} the rank of the skew Vandermonde matrix equals the degree of the minimal
polynomial of the set $\{\alpha_1,\ldots,\alpha_n\}$, which is given by $\lclm(x-\alpha_i\mid i=1,\ldots,n)$.
Hence the rank condition above is equivalent to $\deg(\lclm(x-\alpha_i\mid i=1,\ldots,n))=n$.
In the classical case where $\sigma=\id$, this is equivalent to $\alpha_1,\ldots,\alpha_n$ being distinct,
and the code~$\cE_{\id,\alpha_1,\ldots,\alpha_n}$ is a generalized $[n,k]$-Reed-Solomon code; see \cite[Sec.~5.3]{HP03}.

The proof of Theorem~\ref{T-BoUl14Prop2} follows easily as in classical case of generalized Reed-Solomon codes with the aid of Theorem~\ref{T-VandRank}.
It is well-known that in many cases classical generalized Reed-Solomon codes are cyclic, e.g.\  $\cE_{\id,1,\alpha,\ldots,\alpha^{n-1}}$ is cyclic
if~$\alpha$ is a primitive element of~$\F$ and $n\leq|\F|$.
However, no such statement holds true for skew-polynomial evaluation codes.
Indeed, it is not hard to find examples of codes $\cE_{\sigma,1,\alpha,\ldots,\alpha^{n-1}}$, where~$\alpha$ is a primitive element of~$\F$,
that are not $(\sigma,f)$-skew cyclic for any monic modulus~$f$ of degree~$n$.
The same is true for the evaluation codes $\cE_{\sigma,1,\alpha^{\bracket{1}},\ldots,\alpha^{\bracket{n-1}}}$.

We close this chapter by mentioning that many of the papers cited above also present decoding algorithms for the codes constructed therein.
We refer to the above literature on this important topic.



\printindex
\end{document}